\documentclass[prd,twocolumn,amsmath,amssymb,floatfix,nofootinbib]{revtex4}

\usepackage{amssymb}
\usepackage{amsmath}

\usepackage{sans}

\usepackage{graphicx}
\usepackage{graphics}
\usepackage{dcolumn}
\usepackage{color}
\usepackage{rotate}
\usepackage{fancyhdr}
\usepackage{hyperref}
\usepackage{indentfirst}

\usepackage{umoline}
\usepackage{ulem}

\def\beq{\begin{equation}}
\def\eeq{\end{equation}}
\def\be{\begin{eqnarray}}
\def\ee{\end{eqnarray}}
\def\ba{\begin{eqnarray}}
\def\ea{\end{eqnarray}}
\def\no{\nonumber}

\definecolor{darkred}{rgb}{.743,0,0}

\def\n02b{$0\nu\beta\beta$}
\def\n02bphi{$0\nu\beta\neta\phi$}
\def\lsim{\mathrel{\rlap{\lower4pt\hbox{\hskip1pt$\sim$}}
    \raise1pt\hbox{$<$}}}         
\def\gsim{\mathrel{\rlap{\lower4pt\hbox{\hskip1pt$\sim$}}
    \raise1pt\hbox{$>$}}}         

\begin{document}
\title{Nuclear coalescence from correlation functions}

\author{Kfir Blum}
\email{kfir.blum@cern.ch}
\affiliation{Theory department, CERN, CH-1211 Geneve 23, Switzerland}
\affiliation{Weizmann Institute, Department of Particle Physics and Astrophysics, Rehovot 7610001, Israel}
\author{Masahiro Takimoto}
\affiliation{Weizmann Institute, Department of Particle Physics and Astrophysics, Rehovot 7610001, Israel}

\begin{abstract}
We derive a simple formula relating the cross section for light cluster production (defined via a coalescence factor) to the two-proton correlation function measured in heavy-ion collisions. The formula generalises earlier coalescence-correlation relations found by Scheibl \& Heinz and by Mrowczynski for Gaussian source models. It motivates joint experimental analyses of Hanbury Brown-Twiss (HBT) and cluster yield measurements in existing and future data sets. 
\end{abstract}

\maketitle

\section{Introduction}

The Large Hadron Collider (LHC) made available a diverse data set of production cross sections of light nuclear clusters like deuterons (D), helions ($^3$He) and tritons ($^3$H)~\cite{Adam:2015vda,Acharya:2017fvb}. 
The LHC also brought progress in femtoscopy, the study of the momentum-space correlations of particles emitted in hadronic collisions\footnote{Also known as Hanbury Brown-Twiss (HBT)~\cite{HanburyBrown:1956bqd,Brown:1956zza} analyses.}~\cite{Aamodt:2010jj,Abelev:2012sq,Abelev:2013pqa,Kisiel:2014upa,Adam:2015vja,Adam:2015pya,Szymanski:2016xia,Acharya:2018gyz}. 
These measurements are a  source of information on the state produced in heavy-ion collisions~\cite{Sato:1981ez,Mrowczynski:1987oid,Danielewicz:1992pei,Llope:1995zz,Scheibl:1998tk,Lednicky:2005tb,Mrowczynski:2016xqm,Bellini:2018epz,Braun-Munzinger:2018hat}. 
A review of future prospects can be found in~\cite{Citron:2018lsq}.

In this paper we consider an interesting feature in the data~\cite{Blum:2017qnn}: the anti-correlation between the source homogeniety volume, probed in femtoscopy, and the coalescence factor of nuclear clusters. This correlation was predicted two decades ago in a seminal work by Scheibl \& Heinz~\cite{Scheibl:1998tk}. For a cluster with mass number $A$ and spin $J_A$, observed at vanishing transverse momentum $p_t=0$ in the collider frame, it is summarised by the relation~\cite{Blum:2017qnn,Bellini:2018epz}\footnote{See also~\cite{Llope:1995zz}.}\footnote{See, e.g.~\cite{Mekjian:1977ei,Mekjian:1978us,DasGupta:1981xx} for the appearance of a similar formula within a thermodynamic model.}:
\be
\label{eq:B2ScheiblHeinz}
\frac{\mathcal{B}_A}{m^{2(A-1)}}&\approx&\frac{2J_A+1}{2^A\,\sqrt{A}}\left(\frac{m\,R}{\sqrt{2\pi}}\right)^{3(1-A)}.
\ee
Here, the coalescence factor is defined as 
%
$\mathcal{B}_A=\left(P_A^0\frac{dN_A}{d^3P_A}\right)/\left(p^0\frac{dN}{d^3p}\right)^A$,   
%
where $p^0dN/d^3p$ is the Lorentz-invariant differential yield for constituent nucleons at $p=P_A/A$. The homogeniety volume is parametrised by the HBT radius $R$~\cite{Aamodt:2010jj,Abelev:2012sq,Abelev:2013pqa,Kisiel:2014upa,Adam:2015vja,Adam:2015pya,Szymanski:2016xia,Acharya:2018gyz}\footnote{More practical details about the definition of $R$ are given in Sec.~\ref{s:data}.}. $m\approx0.94$~GeV is the nucleon mass. 

Eq.~(\ref{eq:B2ScheiblHeinz}) was predicted to apply in the limit that the size parameter $d_A$ of the cluster's wave function can be neglected compared to the source homogeniety radius: $d_A\ll R$. For small systems with $R\lesssim d_A$, Eq.~(\ref{eq:B2ScheiblHeinz}) receives a correction via $R^2\to R^2+(d_A/2)^2$. 
At finite $p_t$, Ref.~\cite{Scheibl:1998tk} suggested that Eq.~(\ref{eq:B2ScheiblHeinz}) should be modified by $m\to m_t=\sqrt{m^2+p_t^2}$. 

A comparison of Eq.~(\ref{eq:B2ScheiblHeinz}) to LHC data was presented in Ref.~\cite{Blum:2017qnn}, which used it to extrapolate measurements in Pb-Pb collisions into a prediction of the coalescence factor of D, $^3$He and $^3$H in p-p collisions. This extrapolation is nontrivial. The HBT radius characterising Pb-Pb collisions is $R\sim4$~fm, compared to $R\sim1$~fm measured in p-p collisions. Thus,  Eq.~(\ref{eq:B2ScheiblHeinz}) predicts a large increase in $\mathcal{B}_A$ going from Pb-Pb to p-p: $\mathcal{B}_3^{\rm p-p}/\mathcal{B}_3^{\rm Pb-Pb}\sim4\times10^3$. Subsequent ALICE measurements~\cite{Acharya:2017fvb} in p-p collisions were consistent with this prediction: 
Eq.~(\ref{eq:B2ScheiblHeinz}) appears to work, at least to $\mathcal{O}(1)$ accuracy, over orders of magnitude in $\mathcal{B}_A$. The question we ask (and answer) in this study is, why does it work?

To substantiate this question, note that Ref.~\cite{Scheibl:1998tk} derived Eq.~(\ref{eq:B2ScheiblHeinz}) using a number of assumptions and approximations. A simple source model was used to describe the emission of particles produced in hadronic collisions. This model implemented collective flow with a specific velocity profile and a Gaussian density profile, limited to radial symmetry in the transverse direction. Using a saddle point approximation to evaluate Cooper-Frye  integrals~\cite{Cooper:1974mv}, Ref.~\cite{Scheibl:1998tk} compared their analytic results to a parallel analysis that used the same assumptions to calculate HBT parameters~\cite{Chapman:1995nz}, and found Eq.~(\ref{eq:B2ScheiblHeinz}).

Given this procedure, it is natural to question the theoretical basis for Eq.~(\ref{eq:B2ScheiblHeinz}). For example, as noted in~\cite{Scheibl:1998tk}, it is unlikely that the source model adopted there can actually describe systems ranging from Pb-Pb to p-p in detail. Why then does Eq.~(\ref{eq:B2ScheiblHeinz}) work? can we expect it to remain valid at $p_t>0$; at intermediate centrality; and so on?

The outline of our analysis and main results is as follows. 
In Sec.~\ref{s:QM} we focus on D formation (Sec.~\ref{ss:D}) and two-proton correlations (Sec.~\ref{ss:pair}).  Using non-relativistic quantum mechanics (QM) considerations, in idealistic settings ignoring final-state interactions and other real-life complications, we derive a relation between D formation and two-particle spectra. In Sec.~\ref{ss:BfromC} we extend our results to a relativistic formulation. Our main result is Eq.~(\ref{eq:B2C2}), giving $\mathcal{B}_2$ as an integral of the two-particle correlation function weighted by the D probability density. The derivation does not require a detailed model of the particle emission source. In particular, we need not invoke the assumptions and approximations of~\cite{Chapman:1995nz,Scheibl:1998tk}. Another derivation is shown in App.~\ref{app:kinetic}. 

In Sec.~\ref{s:comp} we show that adopting the same assumptions our formalism reproduces Eq.~(\ref{eq:B2ScheiblHeinz}) as found in~\cite{Scheibl:1998tk}\footnote{\label{fn:1}Apart from the fact that the natural definition we find for $R$ is in the so-called pair rest frame, compared to the longitudinal frame adopted in~\cite{Scheibl:1998tk}, and apart from the replacement $m\to m_t$. Please see Sec.~\ref{s:comp} for details.
}. 
The up shot is that our work makes Eq.~(\ref{eq:B2ScheiblHeinz}) a generic  prediction. If, above, we argued that the model dependence in~\cite{Scheibl:1998tk} makes it a surprise that Eq.~(\ref{eq:B2ScheiblHeinz})  successfully describes systems from Pb-Pb  to p-p, then in light of the discussion in Sec.~\ref{s:QM} it becomes nontrivial to imagine a system for which Eq.~(\ref{eq:B2ScheiblHeinz}) would fail. 
The down side is that Eq.~(\ref{eq:B2ScheiblHeinz}) is essentially a kinematical relation and can teach us relatively little about the dynamics of the state produced in heavy-ion collisions. 
Our analysis bears a connection to (being a less sophisticated version of) Ref.~\cite{Lednicky:2005tb}, which showed that the number of pion pairs produced in Coulomb bound states is related to the number of free pion pairs at small relative momentum. Our work is also close in spirit to work by Mrowczynski~\cite{Mrowczynski:1987oid,Mrowczynski:1989jd,Mrowczynski:1992gc,Mrowczynski:1993cx,Mrowczynski:1994rn,Maj:2004tb}.  

In Sec.~\ref{s:data} we consider complications including final-state interactions and source chaoticity (Sec.~\ref{ss:lam}). We do not address these complications in detail, but show how experimental analyses that take these issues into account can be used to test the coalescence-correlation relation at the cost of some model-dependence. In Sec.~\ref{s:highA} we generalise our results to $A\geq2$, postponing some details to App.~\ref{app:wave}. 
In Sec.~\ref{ss:data} we compare our theoretical results to data. In Sec.~\ref{ss:BAvsR} we recap the results of Ref.~\cite{Blum:2017qnn}, comparing the coalescence-correlation relation with data across systems. 
While our results are consistent with available measurements, the uncertainties are large. Existing experimental analyses were not geared for a direct comparison of femtoscopy and cluster yields. 
This lack motivates dedicated experimental work. 

We conclude in Sec.~\ref{s:sum}.

\section{QM considerations}\label{s:QM}

Hadronic collisions produce a high-excitation state (HXS), characterised by a density matrix $\hat\rho_{\rm HX}$. QM allows to calculate the probability density to find a certain non-relativistic state in the HXS by projecting that state onto $\hat\rho_{\rm HX}$. In this section we use the QM  formalism to derive a relation between D and two-particle spectra. We then convert to Lorentz-invariant quantities.

We emphasise that the QM formulation we use is far from new. It had been utilised in different guises in many early studies including (as a partial list) Refs.~\cite{Sato:1981ez,Llope:1995zz,Koonin:1977fh,Mrowczynski:1987oid,Pratt:1990zq,Mrowczynski:1994rn,Maj:2004tb,Chapman:1995nz,Scheibl:1998tk,Mrowczynski:2016xqm}. Our discussion in Secs.~\ref{ss:D} and~\ref{ss:pair} is merely intended to review the derivation of D and particle pair formation, respectively, in the HXS, recalling that the two phenomenae stem from building blocks that are closely related on general grounds. Our next step, in Sec.~\ref{ss:BfromC}, is to explicitly combine the expressions into a direct relation between coalescence and pair spectra, summarised in Eq.~(\ref{eq:B2C2}). This result, as far as we know, is new to the current work.

\subsection{Deuteron formation}\label{ss:D}
A D at lab-frame momentum $P_d$ is a two-particle (neutron-proton) bound state $|\psi_{P_d}\rangle$ with wave function 
\be
\psi_{P_d}(x_1,x_2)&=&e^{i\vec P_d\vec X}\phi_d(\vec r),\ee
where 
\be 
&\vec X=(\vec x_1+\vec x_2)/2,&\vec r=\vec x_1-\vec x_2\label{eq:Xr}
\ee  
and $\int d^3r|\phi_d(\vec r)|^2=1$. 
The probability density of D in the HXS is~\cite{Sato:1981ez} 
\be\label{eq:d1} \frac{dN_d}{d^3P_d}&=&(2\pi)^{-3}\langle\psi_{P_d}|\hat\rho_{\rm HX}|\psi_{P_d}\rangle\\
&=&\frac{G_d}{(2\pi)^3}\int d^3x_1\int d^3x_2\int d^3x'_1\int d^3x'_2\no\\
&&\psi^*_{P_d}(x'_1,x'_2)\,\psi_{P_d}(x_1,x_2)\,\rho_{\rm 2}\left(x'_1,x'_2;x_1,x_2;t_f\right),\no\ee
where $\rho_{\rm 2}\left(x'_1,x'_2;x_1,x_2;t_f\right)$ is the two-particle reduced HXS density matrix. $G_d$ is a dimensionless normalisation factor. In this section, for simplicity, we assume the existence of a well-defined freeze-out time $t_f$ and consider the HXS density matrix as being specified at the moment $t_f$. We emphasise that this simplification is not essential for the derivation, and our main result [Eq.~(\ref{eq:B2C2}) below] holds also if we allow a finite-duration freeze-out window. An alternative derivation that makes this point manifest is given in App.~\ref{app:kinetic}. 


It is commonly assumed that the HXS density matrix can be factorised into 1-particle density matrices, 
\be\label{eq:rhofct}\rho_{\rm 2}\left(x'_1,x'_2;x_1,x_2;t\right)&=&\rho_{\rm 1}\left(x'_1,x_1;t\right)\rho_{\rm 1}\left(x'_2,x_2;t\right),
\ee
that can in turn be described in terms of Wigner densities $f_1^W$,
\be\label{eq:fW}
\rho_1(x,x';t)&=&\int\frac{d^3k}{(2\pi)^3}e^{i\vec k\left(\vec x'-\vec x\right)}f_1^W\left(\vec k,\frac{\vec x+\vec x'}{2};t\right).
\ee
Inserting Eqs.~(\ref{eq:rhofct}) and~(\ref{eq:fW}) into Eq.~(\ref{eq:d1}) 
we obtain
%
%
\be\label{eq:d20} \frac{dN_d}{d^3P_d}&=&\frac{G_d}{(2\pi)^3}\int d^3R \int\frac{d^3q}{(2\pi)^3}\int d^3r\,\mathcal{D}_d\left(\vec q,\vec r\right)\,\times\\
&&f_1^W\left(\frac{\vec P_d}{2}+\vec q,\vec R +\frac{\vec r}{2};t_f\right)f_1^W\left(\frac{\vec P_d}{2}-\vec q,\vec R -\frac{\vec r}{2};t_f\right),\no\ee
%
where $\mathcal{D}_d$ is the Wigner density of the D,
\be\mathcal{D}_d\left(\vec q,\vec r\right)&=&\int d^3\zeta\,e^{-i\vec q\vec\zeta}\,\phi_d\left(\vec r+\frac{\vec\zeta}{2}\right)\,\phi_d^*\left(\vec r-\frac{\vec\zeta}{2}\right).\ee
In terms of the original variables of Eq.~(\ref{eq:d1}), $\vec R =(\vec x_1+\vec x_1'+\vec x_2+\vec x_2')/4$ is the classical centre of mass coordinate of the two-nucleon system and $\vec r=(\vec x_1+\vec x_1')/2-(\vec x_2+\vec x_2')/2$ is the classical relative coordinate between the nucleons. 

It can be shown that neglecting $\pm\vec q$ inside of the $f_1^W$ functions in Eq.~(\ref{eq:d20}) is a reasonable approximation, valid to $\sim10\%$ accuracy for Pb-Pb collisions~\cite{Scheibl:1998tk}. 
With this approximation we can perform the $q$ integration which gives $\int d^3q\mathcal{D}_d\left(\vec q,\vec r\right)=(2\pi)^3|\phi_d(\vec r)|^2$. 
Defining
\be\label{eq:D(k)}\left|\phi_d\left(\vec r\right)\right|^2&=&\int \frac{d^3k}{(2\pi)^3}\,e^{i\vec k\vec r}\,\mathcal{D}\left(\vec k\right)\ee
we obtain
\be\label{eq:d2}\frac{dN_d}{d^3P_d}&\approx&\frac{G_d}{(2\pi)^6}\int d^3q\,\mathcal{D}\left(\vec q\right)\int d^3R \int d^3r\,e^{i\vec q\vec r}\,\times\\
&&f_1^W\left(\frac{\vec P_d}{2},\vec R +\frac{\vec r}{2};t_f\right)\,f_1^W\left(\frac{\vec P_d}{2},\vec R -\frac{\vec r}{2};t_f\right).\no\ee

Eq.~(\ref{eq:d2}) expresses a non-relativistic QM calculation of the Lorentz non-invariant quantity $dN/d^3P_d$. In Sec.~\ref{ss:BfromC} we return to the problem of connecting this result to the total Lorentz-invariant D yield obtained by integrating over different emission regions in an expanding HXS ``fireball".

\subsection{Nucleon pair emission}\label{ss:pair}
Consider a state $|\psi^s_{p_1,p_2}\rangle$ describing two free propagating protons in a spin-symmetric configuration. Ignoring final-state interactions (FSI), the position space representation of $|\psi^s_{p_1,p_2}\rangle$ is an antisymmetric function of the particle coordinates,
\be
\psi^s_{p_1,p_2}(x_1,x_2)&=&\frac{1}{\sqrt{2}}e^{2i\vec P\vec X}\left(e^{i\vec q\vec r/2}-e^{-i\vec q\vec r/2}\right),\no\\&&
\ee
where the average pair momentum and the momentum difference are defined as 
\be 
&\vec P=\left(\vec p_1+\vec p_2\right)/2,\;\;&\vec q=\vec p_1-\vec p_2.
\ee 
The probability density associated with $|\psi^s_{p_1,p_2}\rangle$ can be calculated as~\cite{Koonin:1977fh,Pratt:1990zq}
%
\be\!\!\!\!\!\!\!\label{eq:2pt1}\frac{dN^s}{d^3p_1d^3p_2}&=&(2\pi)^{-6}\langle\psi^s_{p_1,p_2}|\hat\rho_{\rm HX}|\psi^s_{p_1,p_2}\rangle\\
&=&\frac{G^s_2}{(2\pi)^6}\int d^3x_1\int d^3x_2\int d^3x'_1\int d^3x'_2\no\\
&&\psi^{s*}_{p_1,p_2}(x'_1,x'_2)\,\psi^s_{p_1,p_2}(x_1,x_2)\,\rho_{\rm 2}\left(x'_1,x'_2;x_1,x_2;t_f\right).\no\ee

Assuming unpolarised isospin-invariant HXS, we use the same $\rho_{\rm 2}\left(x'_1,x'_2;x_1,x_2;t_f\right)$ for the proton-proton and proton-neutron reduced density matrix, appearing in Eqs.~(\ref{eq:2pt1}) and~(\ref{eq:d1}). $G^s_2$ is a normalisation constant. 
Inserting Eqs.~(\ref{eq:rhofct}) and~(\ref{eq:fW}) into Eq.~(\ref{eq:2pt1}) 
we obtain
\be\label{eq:Unit}\frac{dN^s}{d^3p_1d^3p_2}&=&G_2^s\left(\mathcal{A}_2\left(p_1,p_2\right)-\mathcal{F}_2\left(P,q\right)\right),\\
\label{eq:F2}\mathcal{F}_2\left(P,q\right)&=&\frac{1}{(2\pi)^6}\int d^3R\int d^3r\,e^{i\vec q\vec r}\,\times\no\\
&&f_1^W\left(\vec P,\vec R+\frac{\vec r}{2};t_f\right)\,f_1^W\left(\vec P,\vec R-\frac{\vec r}{2};t_f\right),\no\\
\mathcal{A}_2\left(p_1,p_2\right)&=&\frac{1}{(2\pi)^6}
\int d^3xf_1^W\left(\vec p_1,\vec x;t_f\right)\,\int d^3xf_1^W\left(\vec p_2,\vec x;t_f\right).\no\ee
We could express $\mathcal{A}_2$ in Eq.~(\ref{eq:Unit}) in terms of $P,q$, but we keep $p_1,p_2$ for clarity. The $P,q$ notation is useful for the $\mathcal{F}_2$ term, which expresses the QM correlation. 

We can repeat the same steps above for the spin anti-symmetric state $|\psi^a_{p_1,p_2}\rangle$, for which the wave function is an symmetric function of the particle coordinates. We find
\be\label{eq:Unita}\frac{dN^a}{d^3p_1d^3p_2}&=&G_2^a\left(\mathcal{A}_2\left(p_1,p_2\right)+\mathcal{F}_2\left(P,q\right)\right),\ee
with $G_2^a=G_2^s/3$.

\subsection{Coalescence from two-particle correlations}\label{ss:BfromC}

Eqs.~(\ref{eq:d2}) and~(\ref{eq:Unit}-\ref{eq:Unita}) give the number of D's and proton pairs, respectively, per differential momentum element when all momenta involved are small. The Lorentz-invariant version of the quantities on the LHS of these equations are $\gamma_d\,dN_d/d^3P_d$ and $\gamma_1\gamma_2\,dN^{s,a}/d^3p_1d^3p_2$. Subtleties arise in the computation of the RHS because for a relativistically expanding HXS, different parts of the particle emission region are moving relativistically w.r.t. other parts. This makes the spatial integrations 
nontrivial~\cite{Cooper:1974mv}. In addition, instead of a homogeneous freeze-out time $t_f$ we expect a freeze-out surface $t_f=t_f(\vec R)$. We now consider these issues.

Inspecting Eqs.~(\ref{eq:d2}) and~(\ref{eq:Unit}), we can write a differential coalescence-correlation relation
\be\label{eq:d3qm} 
\frac{d}{d^3R}\left(\frac{dN_d}{d^3P_d}\right)
&\approx&G_d\frac{d}{d^3R}\int d^3q\,\mathcal{D}(\vec q)\,
\mathcal{F}_2\left(\frac{\vec P_d}{2},\vec q\right).\no\\&&
\ee
The differential presentation reveals model-independence in terms of the details of freeze-out. By either  plugging-in Eq.~(\ref{eq:D(k)}), or proceeding directly from Eq.~(\ref{eq:d2}), we have
%
\be\label{eq:d20L1} \frac{d}{d^3R}\left(\frac{dN_d}{d^3P_d}\right)&=&\frac{G_d}{(2\pi)^3}\,f_1^W\left(\frac{\vec P_d}{2},\vec R;t_f\right)\,\times\no\\
&&\int d^3r\left|\phi_d(\vec r)\right|^2f_1^W\left(\frac{\vec P_d}{2},\vec R-\vec r;t_f\right).\no\\&&\ee
It is natural to regard the RHS of Eq.~(\ref{eq:d20L1}) as a Lorentz-invariant distribution function $f_d$. 
This was done in Ref.~\cite{Scheibl:1998tk}, which used the Cooper-Frye prescription~\cite{Cooper:1974mv} to make the replacement $\gamma_d\int d^3R\,f_d\to(1/2m)\int \left[d^3\sigma_\mu P_d^\mu\right]f_d$, where $d^3\sigma^\mu$ is the volume element perpendicular to the HXS relativistic freeze-out surface. 

While Ref.~\cite{Scheibl:1998tk} (which focused on D formation) arrived at this procedure directly from Eq.~(\ref{eq:d2}), the same implementation of freeze-out w.r.t. the integration over centre of mass coordinate $\vec R$ can be used in integrating the coalescence-correlation relation expressed by Eq.~(\ref{eq:d3qm}).  
There is no need to specify the details of the freeze-out surface $t_f(\vec R)$ 
because Eq.~(\ref{eq:d3qm}) relates the pair emissivity and the D emissivity per differential volume element $d^3R$ in the HXS. Having noted this point, we can drop the differential $d^3R$ in Eq.~(\ref{eq:d3qm}) and consider it as a relation between total D and pair yields.

Let us now make contact with measurements. Experimental collaborations report the (Lorentz-invariant) coalescence factor 
\be\label{eq:B2}\mathcal{B}_2(p)&=&\frac{P_d^0\,\frac{dN_d}{d^3P_d}}{\left(p^0\frac{dN}{d^3p}\right)^2},\ee
with $p=P_d/2$ and where $p^0\frac{dN}{d^3p}$ is the unpolarised proton yield. The two-particle correlation function is constructed as
\be \label{eq:C2base}C_2(P,q)&=&\frac{p_1^0\,p_2^0\frac{dN}{d^3p_1d^3p_2}}{\left(p_1^0\frac{dN}{d^3p_1}\right)\left(p_2^0\frac{dN}{d^3p_2}\right)}.
\ee
The numerator on the RHS of Eq.~(\ref{eq:C2base}) sums together the different spin states of the proton pair. 
In the denominator, the unpolarised differential yields at $p_1$ and $p_2$ are obtained by scrambling between proton pairs from different events. 
 
Still provisionally neglecting FSI and other complications (which would be discussed later), Ref.~\cite{Adam:2015vja} parametrised two-proton correlation measurements in a way that can be put as
\be\label{eq:C2}C_2(P,q)&=&1-\frac{G_2^s-G_2^a}{G_2^s+G_2^a}\,\mathcal{C}_2(P,q).\ee %
By examining the $q$ dependence we see that the $\mathcal{C}_2$ term in Eq.~(\ref{eq:C2}) comes from the $\mathcal{F}_2$ term in Eq.~(\ref{eq:Unit}), while the 1 comes from the $\mathcal{A}_2$ term there. More precisely, in the non-relativistic limit we have
\be \mathcal{C}^{\rm PRF}_2\left(|\vec q|\ll m\right)&=&\frac{\mathcal{F}_2}{\mathcal{A}_2},\ee
where the superscript PRF instructs us that $q$ in $\mathcal{C}_2^{\rm PRF}$ is defined in the pair centre of mass frame. 
In the same limit, Eqs.~(\ref{eq:d3qm}) and~(\ref{eq:B2}) show that
\be\mathcal{B}_2(p)&=&\frac{G_d}{G_2^s+G_2^a}\frac{2\,m}{m^2\mathcal{A}_2}\int d^3q\,\mathcal{D}(\vec q)\,\mathcal{F}_2(\vec p,\vec q).\ee
%
Assuming unpolarised isospin-symmetric HXS~\cite{Mattiello:1996gq} 
we have
\be\frac{G_d}{G^s_2+G_2^a}&=&\frac{3}{3+1}.
\ee
%
%
Using these conventions and noting that $\gamma_1\approx\gamma_2\approx\gamma_d$ for small $|\vec q|\ll m$, we are finally led to the result:
\be\label{eq:B2C2}\boxed{\;\mathcal{B}_2(p)\;\approx\;\frac{3}{2\,m}\int d^3q\,\mathcal{D}(\vec q)\,\mathcal{C}^{\rm PRF}_2\left(\vec p,\vec q\right).\;}\ee
%
Following the discussion around Eq.~(\ref{eq:d3qm}), this result is not limited to non-relativistic $p$. It is limited to non-relativistic $|\vec q|^2\ll m^2$, but that is not a real concern because both $\mathcal{C}_2$ and $\mathcal{D}$ cut-off at $|\vec q|\sim0.1\,m$.

We comment that the coalescence factor $\mathcal{B}_2(p)$ is defined for on-shell D with $P_d^2=4p^2\approx(2m)^2$. Thus, there will actually be no on-shell proton pairs that satisfy $p_1^2=p_2^2=m^2$ along with $(p_1+p_2)/2=p$ at $q\neq0$. This problem comes from neglecting  corrections of order $\vec q^2/m^2$ in the derivation of Eq.~(\ref{eq:B2C2}). We can find on-shell proton pairs to construct $\mathcal{C}_2^{\rm PRF}$ by allowing the energy component $P^0$ of the $P$ 4-vector in Eq.~(\ref{eq:C2base}) to deviate from $p^0$ of Eq.~(\ref{eq:B2}), while at the same time enforcing $\vec P=\vec P_d/2=\vec p$. In other words, we let $p$ on the LHS of Eq.~(\ref{eq:B2C2}) denote the 4-momentum per nucleon of the on-shell D, and we equate $\vec p$ between the LHS and the RHS, but we do not enforce $p^0$ on the RHS to match $p^0$ on the LHS. Corrections due to this approximation are of order $\vec q^2/m^2$. 

\section{Comparison with previous work}\label{s:comp}

Scheibl \& Heinz~\cite{Scheibl:1998tk} used a Gaussian source model (GSM) of the HXS 1-particle Wigner densities to calculate coalescence and two-particle correlations (following~\cite{Chapman:1995nz} on the latter), and expressed the coalescence factor in terms of the HBT radius parameters computed in their model. To obtain analytic expressions, the D wave function was taken to be Gaussian,
\be\label{ref:dGauss}\phi_d(\vec r)&=&\frac{e^{-\frac{\vec r^2}{2d^2}}}{\left(\pi d^2\right)^{\frac{3}{4}}}\ee
with $d=3.2$~fm. 
This leads to
\be\mathcal{D}(\vec k)&=&e^{-\frac{\vec k^2d^2}{4}}.\ee
For the HBT analysis, Ref.~\cite{Scheibl:1998tk} used the parameters 
$R_{\perp}$ and $R_{||}$ in terms of which the correlation function in their model is given by
\be\label{eq:C2gauss}\mathcal{C}_2^{\rm PRF}&=&e^{-R_{\perp}^2\vec q^2_{\perp}-R^2_{||}\vec q^2_l},\;\;\;\;\;\;{\rm {\bf(GSM)}}\ee
where $\vec q_l$ is the component of $\vec q$ parallel to the beam axis and $\vec q_{\perp}$ spans the transverse direction. Plugging these expressions for $\mathcal{D}$ and $\mathcal{C}_2^{\rm PRF}$ in Eq.~(\ref{eq:B2C2}) we find\footnote{See also~\cite{Llope:1995zz,Murray:2000cw}.}
\be\label{eq:B2SP}\mathcal{B}_2&=&\frac{3\pi^{\frac{3}{2}}}{2\, m\left(R_{\perp}^2+\left(\frac{d}{2}\right)^2\right)\sqrt{R_{||}^2+\left(\frac{d}{2}\right)^2}},\;\;\;{\rm {\bf(GSM)}}.\no\\&&\ee
%
This reproduces Eq.~(\ref{eq:B2ScheiblHeinz}) and the main result of~\cite{Scheibl:1998tk} (see Eqs.~(6.3) and~(4.12) there), up to the replacement $m\to m_t=\sqrt{m^2+\vec p_t^2}$. 
Please note that we have defined $R_\perp$ and $R_{||}$ in the PRF, while~\cite{Scheibl:1998tk} defined these parameters in the YKP frame~\cite{Chapman:1995nz,Yano:1978gk,Podgoretsky:1982xu,Wu:1996wk} which is offset by a transverse boost compared to the PRF. 

%
Mrowczynski discussed the connection between coalescence and two-particle correlations in a series of papers~\cite{Mrowczynski:1987oid,Mrowczynski:1989jd,Mrowczynski:1992gc,Mrowczynski:1993cx,Mrowczynski:1994rn,Maj:2004tb}. This program resulted in a QM sum rule of the neutron-proton correlation function, that was proposed to give the D coalescence factor as a $q$-integral on the correlation function~\cite{Maj:2004tb}. The power of this idea was in that there was no need to correct the measured correlation function for long- or short-range final state interactions: the sum rule should apply directly to the observable correlation. In practice, this suggestion fails, apparently because the $q$-integral proposed in~\cite{Maj:2004tb} receives contributions from large-$q$ regions in the integration. 

In comparison to the sum rule of~\cite{Mrowczynski:1994rn,Maj:2004tb}, Eq.~(\ref{eq:B2C2}) is less ambitious. The correlation function entering Eq.~(\ref{eq:B2C2}) does need to be corrected for final state interactions, because it assumes a kinetic picture where an HXS density matrix can be defined and projected into propagating particles. Eq.~(\ref{eq:B2C2}) also invokes assumptions such as isospin symmetry and smoothness for the HXS freezeout surface. In return, however, the RHS of Eq.~(\ref{eq:B2C2}) receives no contributions from large-$q$ modes because $\mathcal{D}(\vec q)$ in the integrand constrains the support to the small $q$ region, $|\vec q|\lesssim0.1m$. 

A QM derivation of the coalescence factor using a specific one-dimensional Gaussian source model was given in Ref.~\cite{Mrowczynski:2016xqm}. This derivation agrees with Eq.~(\ref{eq:B2SP}) up to the replacement $m\to p^0=m\gamma_d$. 

\section{Real-life complications, $A>2$ clusters, and comparing to data}\label{s:data}

Eq.~(\ref{eq:B2C2}) is idealistic. In practice we cannot pull out a directly measured correlation function $\mathcal{C}_2$, plug into Eq.~(\ref{eq:B2C2}) and calculate $\mathcal{B}_2$. Two main complications, preventing direct implementation of Eq.~(\ref{eq:B2C2}), are: (i) Long-lived resonances, decaying outside of the freeze-out surface of the HXS, distort the correlations. (ii) Long-range Coulomb and short-range strong nuclear FSI cause the two-particle wave function to differ from the plane-wave form. For proton pairs, FSI actually dominate the correlation function, meaning that the QM statistics contribution must be extracted indirectly as a sub-leading contribution to the actual observable $\mathcal{C}_2$. To make things more difficult, different spin states exhibit different short-range FSI.

We will not address the complications above in detail in this paper, deferring such refinements to future work. Instead, we build on femtoscopy data analyses that explicitly treat items (i-ii). The price we pay is to introduce model-dependence, that enters via an assumed simple analytic form for the correlation function. Our procedure and results are explained in the next sections.

\subsection{The chaoticity parameter $\lambda$}\label{ss:lam}

The GSM assumed  in~\cite{Chapman:1995nz,Scheibl:1998tk,Mrowczynski:2016xqm} predicts not only the shape, but also the normalisation of $\mathcal{C}_2$: it predicts $\mathcal{C}_2^{\rm PRF}(\vec q\to0)=1$. In reality, measurements  show $\mathcal{C}_2^{\rm PRF}(\vec q\to0)\to\lambda<1$, where $\lambda$ is known as the chaoticity (or intercept) parameter~\cite{Wiedemann:1996ig,Akkelin:2001nd}. In HBT analyses of pions,  $\lambda<1$ follows from the fact that a sizeable fraction of the pions come from the decay of long-lived resonances, leading to a non-Gaussian contribution to $\mathcal{C}_2$ that is concentrated at very small $|\vec q|$ and cannot be resolved experimentally~\cite{Wiedemann:1996ig}. In HBT analyses of proton pairs, hyperons  are the resonant contamination~\cite{Wang:1999xq,Wang:1999bf,Szymanski:2016xia}. 
%
Since strong FSI between $p\Lambda$ and $pp$ are crucial in shaping the $p\Lambda$ and $pp$ correlation functions, studies~\cite{Wang:1999xq,Wang:1999bf,Szymanski:2016xia,Adam:2015vja} separate the $p\Lambda\to pp$ and genuine $pp$ contributions entering the observed $pp$ correlation into different terms, that are fit in a combined analysis. In~\cite{Adam:2015vja,Szymanski:2016xia}, separate chaoticity parameters $\lambda_{pp},\,\lambda_{p\Lambda}$ were assigned to the genuine $pp$ pairs and the pairs coming from $p\Lambda\to pp$. 
The value of $\lambda$ defined in this way could reflect intrinsic departures of the source functions from Gaussianity.

In Ref.~\cite{Chapman:1995nz} (and many other analyses in the literature), $\lambda$ was introduced as a free parameter. Thus, it did not enter into the coalescence-HBT correspondence of Ref.~\cite{Scheibl:1998tk}. However, Eq.~(\ref{eq:B2C2}) shows that $\mathcal{B}_2$ is directly proportional to a $q$-moment of $\mathcal{C}_2^{\rm PRF}$. If we adopt the Gaussian form together with the $\lambda$ modification as an empirical description of $\mathcal{C}_2$, 
\be\label{eq:C2gaussl}\mathcal{C}_2^{\rm PRF}&=&\lambda\,e^{-R_{\perp}^2\vec q^2_{\perp}-R_{||}\vec q^2_l},\,{\rm {\bf(GSM,\,chaoticity\,\lambda)}}\no\\&&\ee
then $\mathcal{B}_2$ should match Eq.~(\ref{eq:B2SP}) simply multiplied by the experimentally deduced value of $\lambda$:
\be\label{eq:B2SPlam}\mathcal{B}_2&=&\frac{3\pi^{\frac{3}{2}}\lambda}{2m\left(R_{\perp}^2+\left(\frac{d}{2}\right)^2\right)\sqrt{R_{||}^2+\left(\frac{d}{2}\right)^2}},\,{\rm {\bf(GSM,\,chaoticity\,\lambda)}}.\no\\&&\ee
%

%

\subsection{$A\geq2$}\label{s:highA}
Eq.~(\ref{eq:B2C2}) can be generalised to clusters with $A\geq2$. Assuming an $(A-1)$-dimensional symmetric Gaussian form for the cluster's relative coordinate wave function, and assuming that the $A$-particle correlation function can be decomposed as a product of 2-particle Gaussian correlators described by the same HBT radii $R_{\perp}$ and $R_{||}$ and chaoticity $\lambda$, then the analogue of Eq.~(\ref{eq:B2SPlam}) is:
%
\be\label{eq:BA}\frac{\mathcal{B}_A}{m^{2(A-1)}}&=&\lambda^{\frac{A}{2}}\frac{2J_A+1}{2^A\sqrt{A}}\,\times\no\\
&&\left[\frac{(2\pi)^{\frac{3}{2}}}{m^3\left(R_{\perp}^2+\left(\frac{d_A}{2}\right)^2\right)\sqrt{R_{||}^2+\left(\frac{d_A}{2}\right)^2}}\right]^{A-1}.\no\\&&\ee
The definition of the cluster wave function and its size parameter $d_A$, used in Eq.~(\ref{eq:BA}), are given in App.~\ref{app:wave}.

%

\subsection{Comparing to data}\label{ss:data}

Experimental collaborations often report the results of HBT analyses in terms of empirical fit parameters $R$ and $\lambda$~\cite{Aamodt:2010jj,Abelev:2012sq,Abelev:2013pqa,Kisiel:2014upa,Adam:2015vja,Adam:2015pya,Szymanski:2016xia,Acharya:2018gyz}, assuming Eq.~(\ref{eq:C2gaussl}) and accounting explicitly for the spin symmetry of the pair wave function and for the distortion due to FSI~\cite{Koonin:1977fh,Lednicky:1981su,Lednicky:2005tb}. 
To compare our theoretical results to data, we will therefore use Eqs.~(\ref{eq:B2SPlam}) and~(\ref{eq:BA}). We further take the extra simplification of a 1-dimensional HBT parametrisation with $R_{\perp}=R_{||}=R$.  

Pion, kaon, and proton femtoscopy results in Pb-Pb collisions were reported in~\cite{Adam:2015vja,Szymanski:2016xia}. Results for proton and kaon femtoscopy in p-p collisions were given in~\cite{Acharya:2018gyz} and~\cite{Abelev:2012sq}, respectively. The kaon results are of potential use because Ref.~\cite{Adam:2015vja} showed compatible results for the parameters $R$ and $\lambda$ obtained in proton and kaon correlations at the same $m_t$. 
Using these HBT analyses we can calculate the RHS of Eqs.~(\ref{eq:B2SPlam}) and~(\ref{eq:BA}), and compare to experimental data on the production of light nuclei~\cite{Adam:2015vda,Acharya:2017fvb}. 

\subsubsection{Pb-Pb collisions}
\begin{figure}[htbp]
  \begin{center}
   \includegraphics[width=0.495\textwidth]{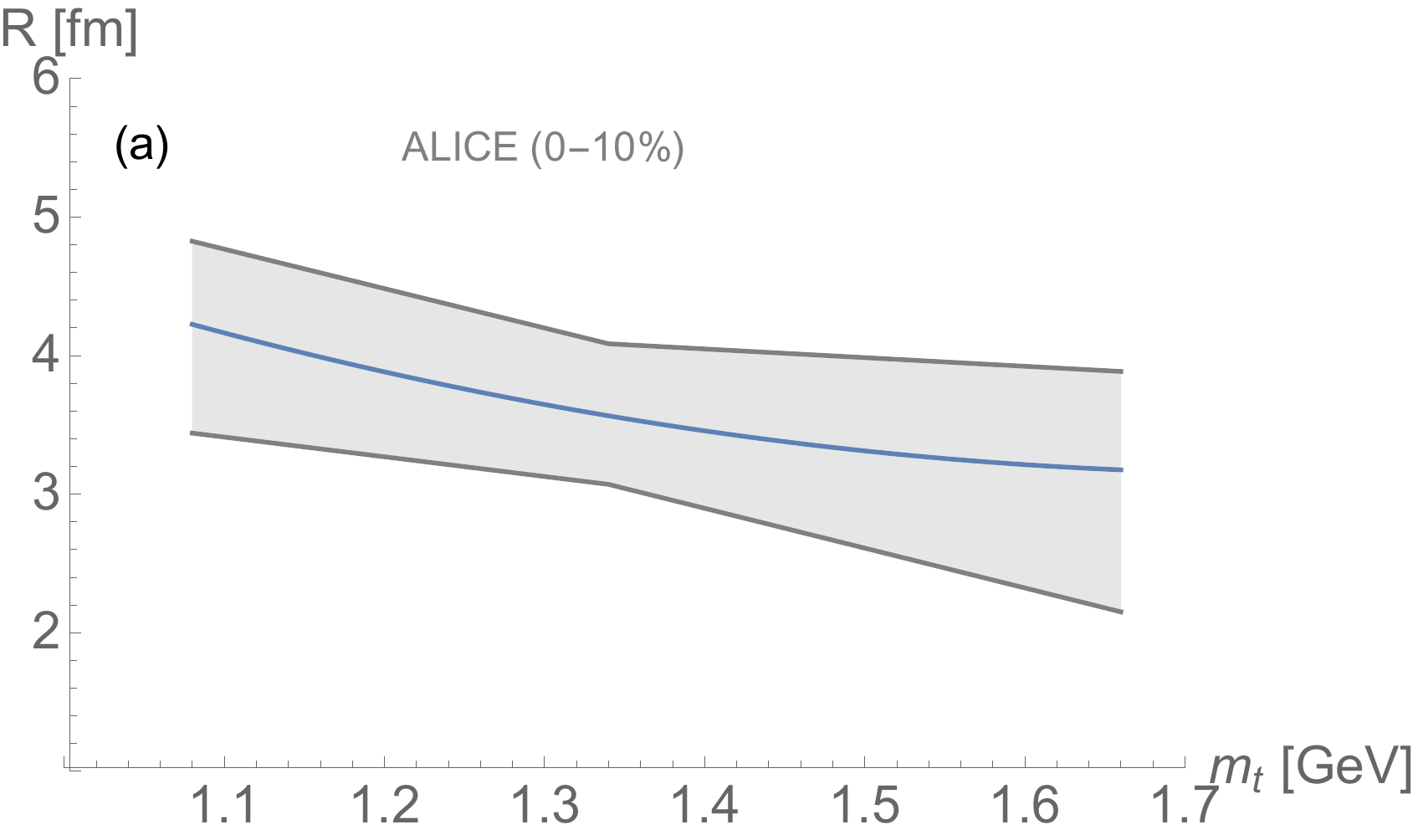}
    \includegraphics[width=0.495\textwidth]{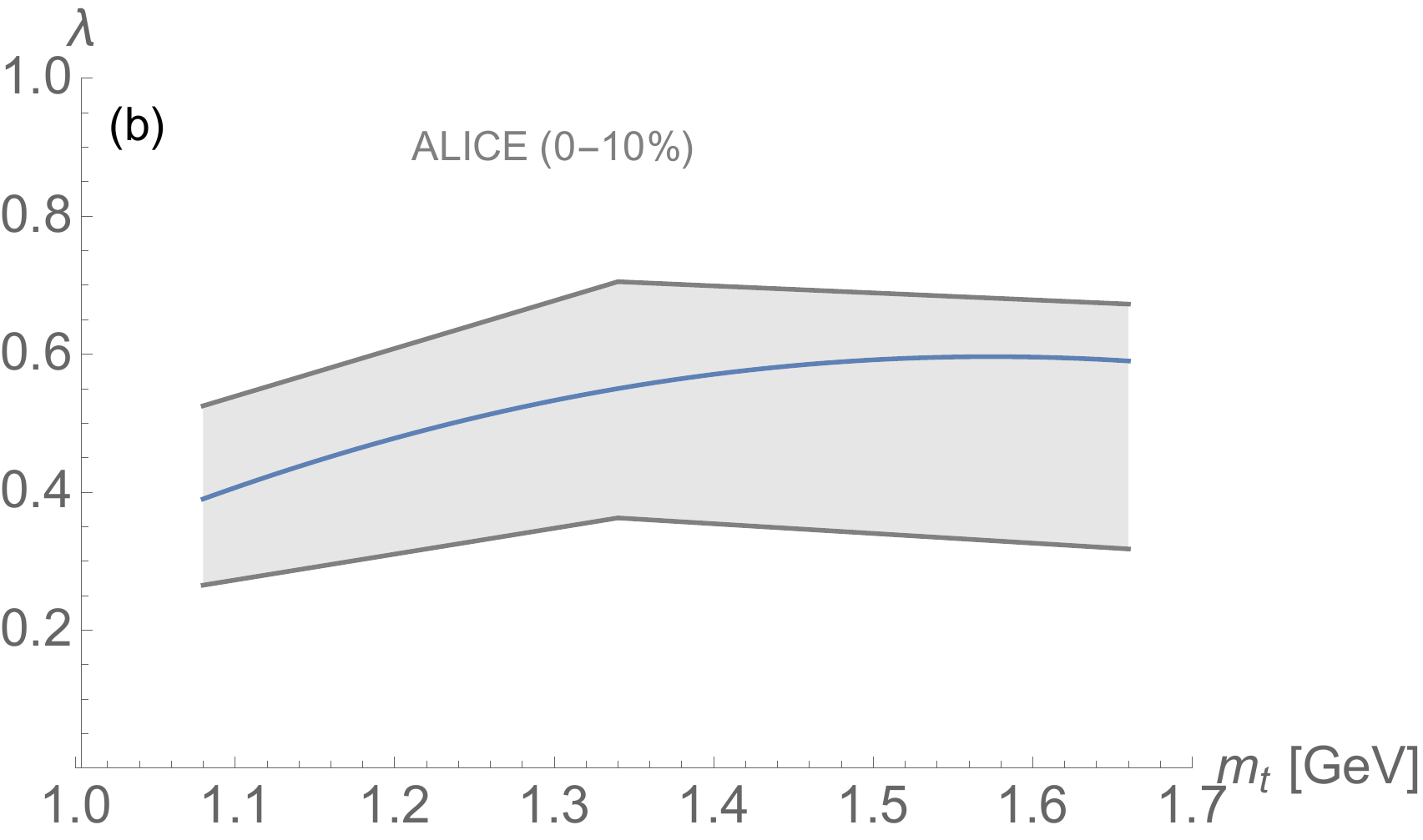}
    \includegraphics[width=0.495\textwidth]{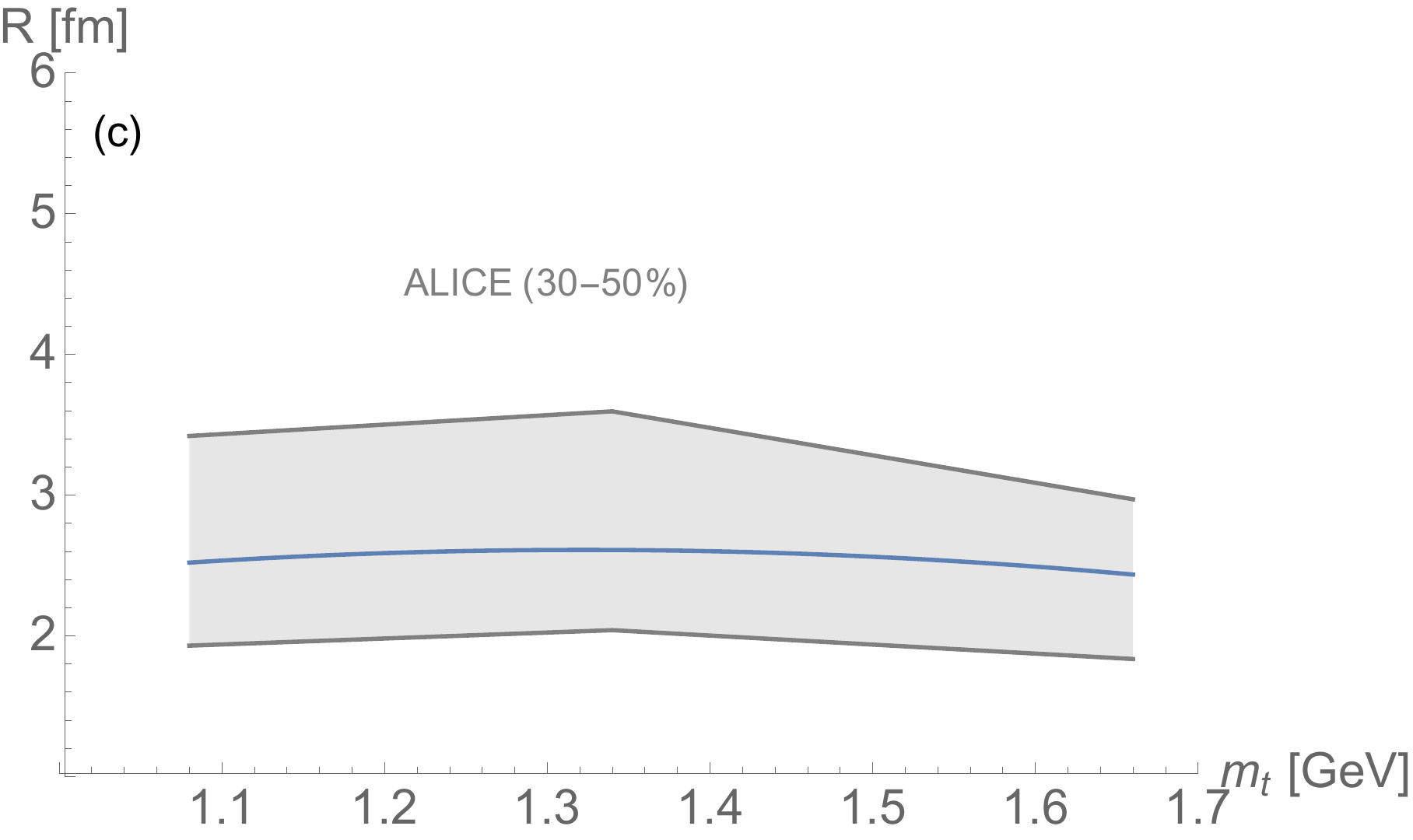}
    \includegraphics[width=0.495\textwidth]{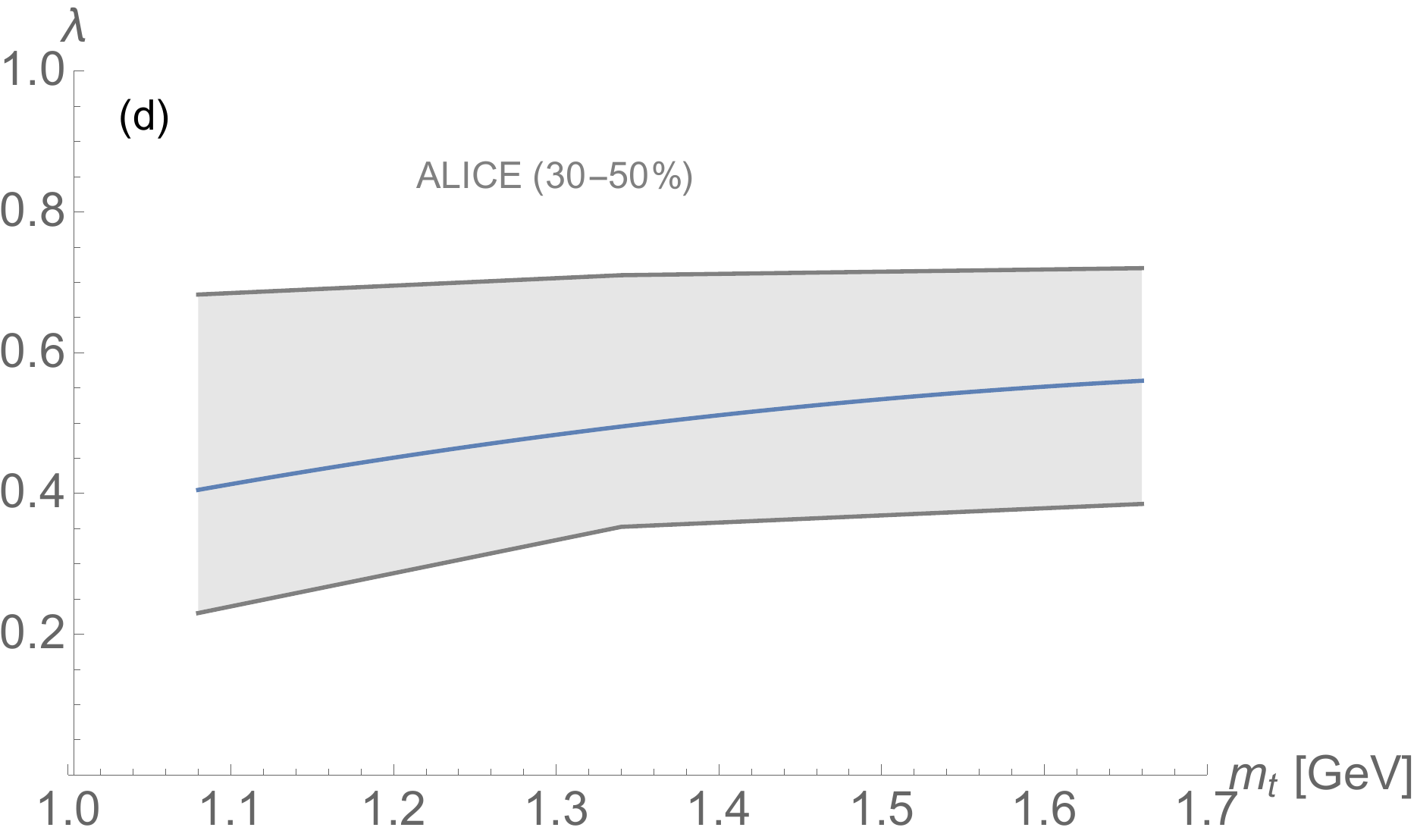}
  \end{center}
  \caption{Experimental fit results for the 1-dimensional HBT radius $R$ and $\lambda$ parameters, extracted from correlations of $pp$, $p\Lambda$, and their anti-particles in central (0-10\%; {\bf panels (a) and (b)}) and intermediate centrality (30-50\%; {\bf panels (c) and (d)}) Pb-Pb collisions at $\sqrt{s}=2.76$~TeV~\cite{Adam:2015vja,Szymanski:2016xia}.}
  \label{fig:RlamPbPb}
\end{figure}

The {\bf top two panels} in Fig.~\ref{fig:RlamPbPb} summarise experimental results for $R$ and $\lambda$ in central (0-10\%) Pb-Pb collisions at $\sqrt{s}=2.76$~TeV~\cite{Adam:2015vja}\footnote{Useful details can be found in Tables 7.4-7.9 in~\cite{Szymanski:2016xia}.}. The {\bf bottom two panels} show $R$ and $\lambda$ obtained in intermediate centrality (30-50\%) data. For $R$, we show the average values found for $pp$ and $\bar p\bar p$ pairs. The uncertainties are mostly systematic, and the width of the band neglects the statistical uncertainty. For $\lambda$, we show the sum $\lambda_{pp}+\lambda_{p\Lambda}$, take the average of the systematic uncertainty, and average the result between particles and anti-particles\footnote{The reason to use the sum of $\lambda_{pp}+\lambda_{p\Lambda}$, and not just $\lambda_{pp}$, is that we are interested to use Eq.~(\ref{eq:B2SPlam}) which assumes the same single-particle spectrum normalisation in the definition of $\mathcal{C}_2$ and $\mathcal{B}_2$. However, the single-particle spectrum entering the denominator of $\mathcal{B}_2$ in the experimental analysis includes only the prompt contribution, while the denominator in the $\mathcal{C}_2$ experimental analysis with $pp$ and $p\Lambda$ terms explicitly separated includes both prompt and secondary protons.}. 

\begin{figure}[htbp]
  \begin{center}
   \includegraphics[width=0.495\textwidth]{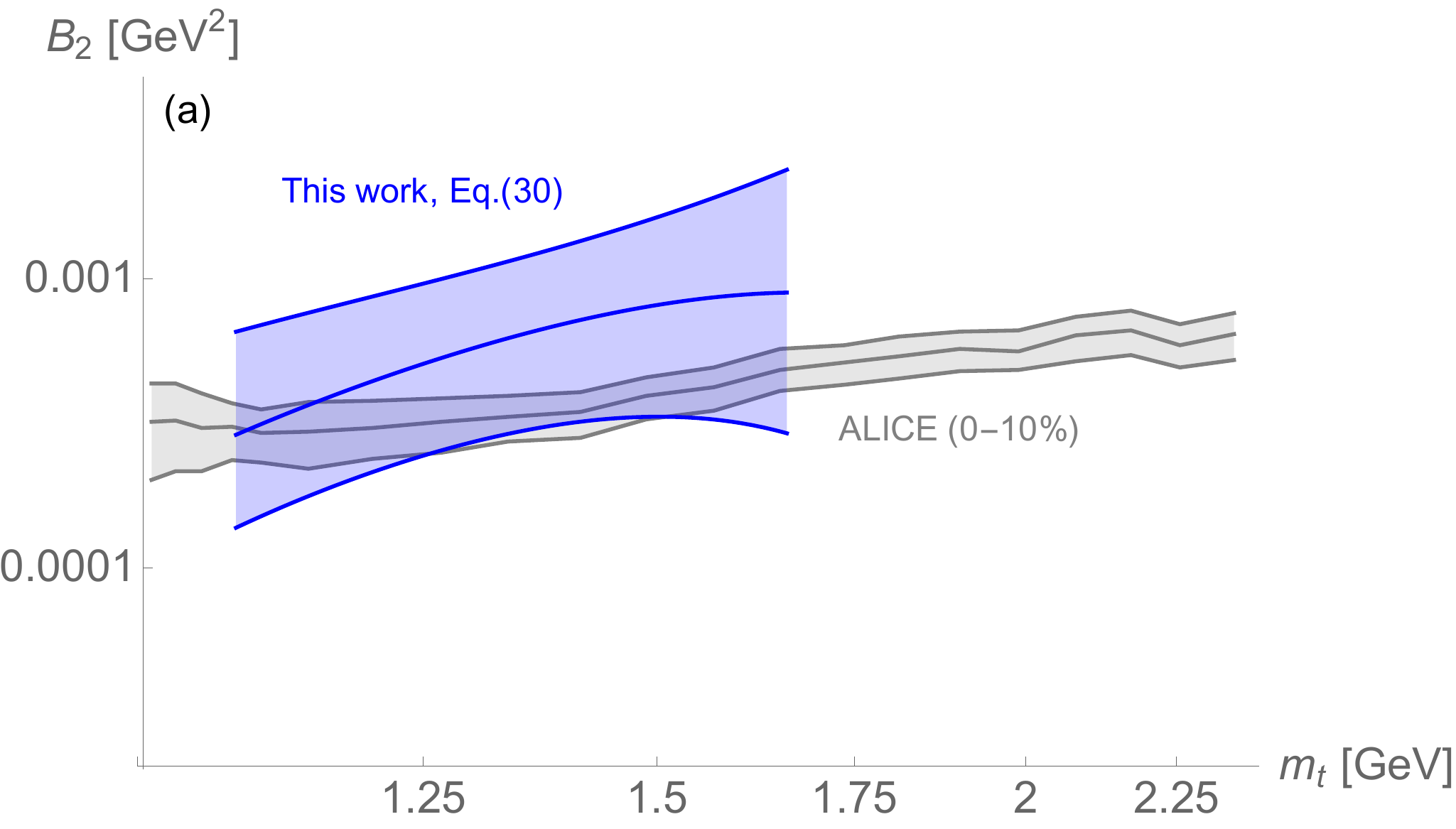}
    \includegraphics[width=0.495\textwidth]{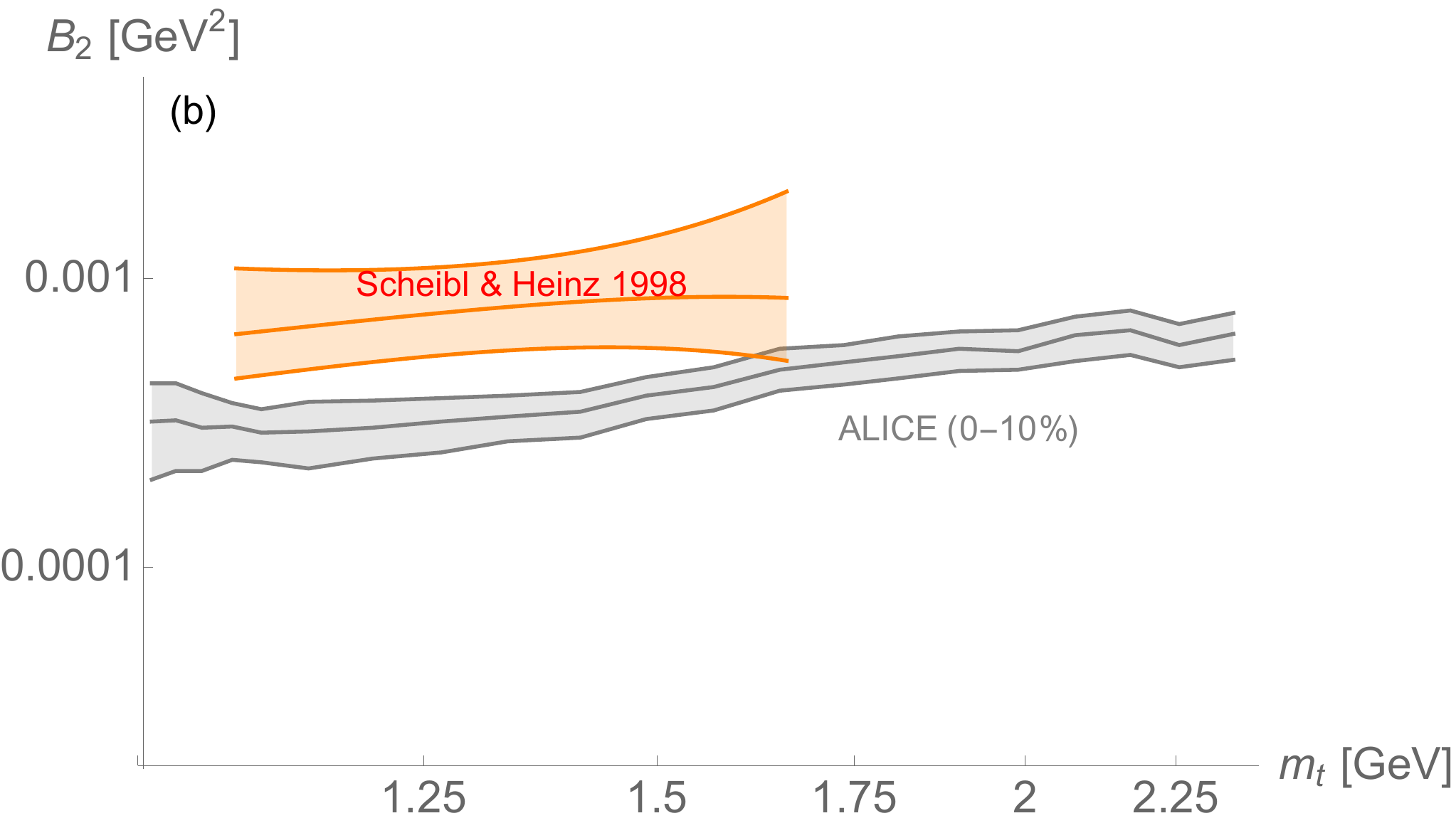}
     \includegraphics[width=0.495\textwidth]{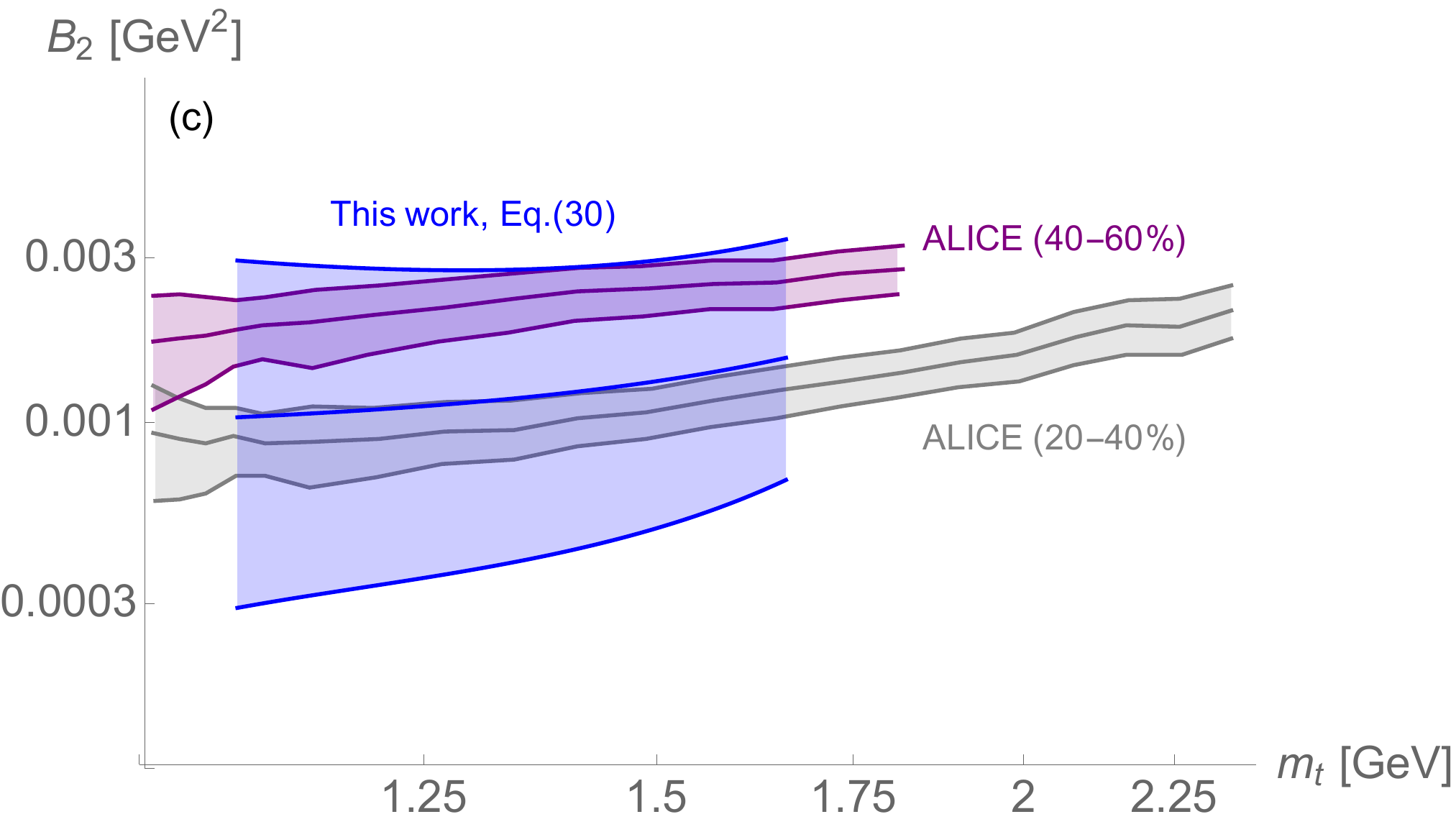}
    \includegraphics[width=0.495\textwidth]{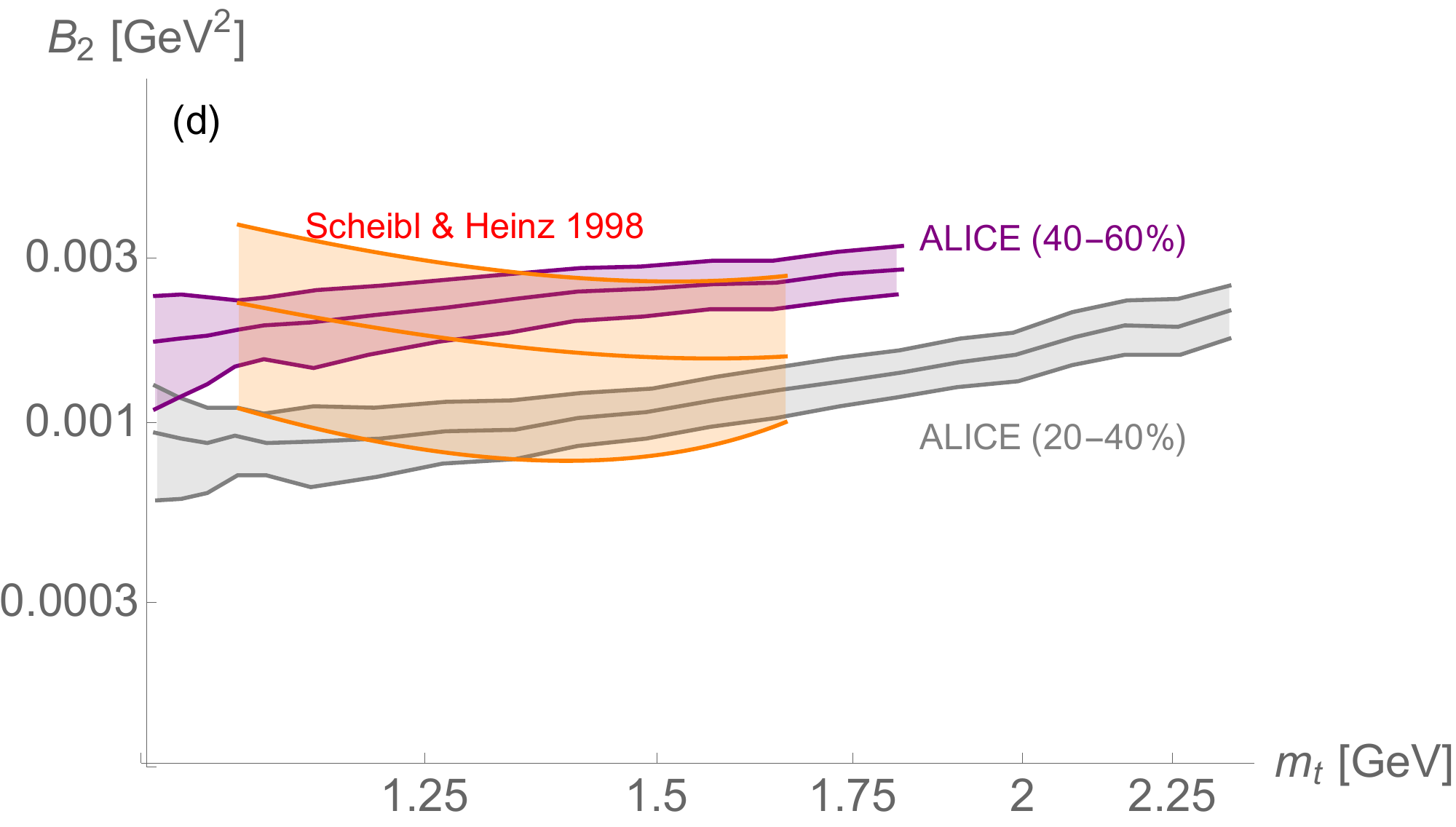}
    \end{center}
  \caption{{\bf Panels (a) and (b):} experimental results for $\mathcal{B}_2$ from central (0-10\%) PbPb collisions at $\sqrt{s}=2.76$~TeV~\cite{Adam:2015vda}, shown by grey band, compared to Eq.~(\ref{eq:B2SPlam}) derived here (blue band) and to the prediction of Ref.~\cite{Scheibl:1998tk} (orange band). The coalescence calculation uses the experimentally extracted HBT $R$ and $\lambda$ parameters shown in Fig.~\ref{fig:RlamPbPb}. {\bf Panels (c) and (d):} experimental values of $\mathcal{B}_2$ from two intermediate centrality classes, (20-40\%) and (40-60\%), and the theoretical prediction calculated using HBT data from events at (30-50\%).}
  \label{fig:B2PbPb}
\end{figure}

Plugging these values of $R,\,\lambda$ into the RHS of Eq.~(\ref{eq:B2SPlam}), we obtain a prediction for $\mathcal{B}_2$. The result for (0-10\%) centrality events is shown by the blue shaded band in the {\bf topmost panel} of Fig.~\ref{fig:B2PbPb}. The uncertainty of the theory prediction was obtained by using the lower value for $\lambda$ and the upper value for $R$ to calculate the lower value of the predicted $\mathcal{B}_2$, and vice-verse. 
An experimental measurement of $\mathcal{B}_2$~\cite{Adam:2015vda} is shown in the same plot as a grey band. We can also compare the data with the theoretical prediction of Ref.~\cite{Scheibl:1998tk}; this is done in the {\bf second from top panel} of Fig.~\ref{fig:B2PbPb}.
In the {\bf bottom two panels} of Fig.~\ref{fig:B2PbPb} we repeat the analysis using the intermediate centrality (30-50\%) HBT parameters, compared to $\mathcal{B}_2$ data corresponding to events at (20-40\%) and (40-60\%) centrality events\footnote{Note that the analysis of Ref.~\cite{Scheibl:1998tk} was restricted to radially symmetric HXS in the plane transverse to the beam axis. It should not, in principle, be valid for intermediate centrality.}. 

\begin{figure}[htbp]
  \begin{center}
   \includegraphics[width=0.495\textwidth]{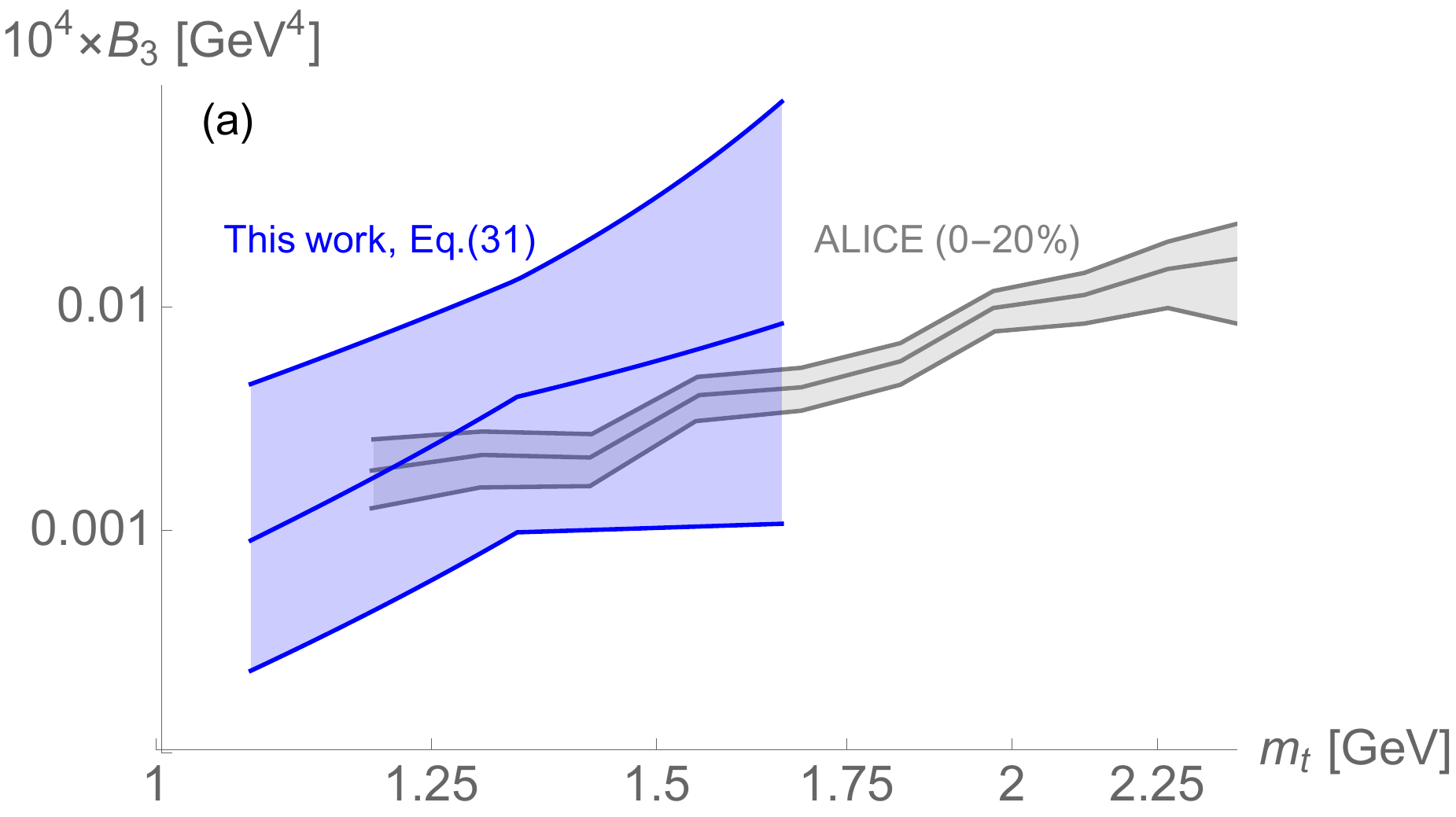}
    \includegraphics[width=0.495\textwidth]{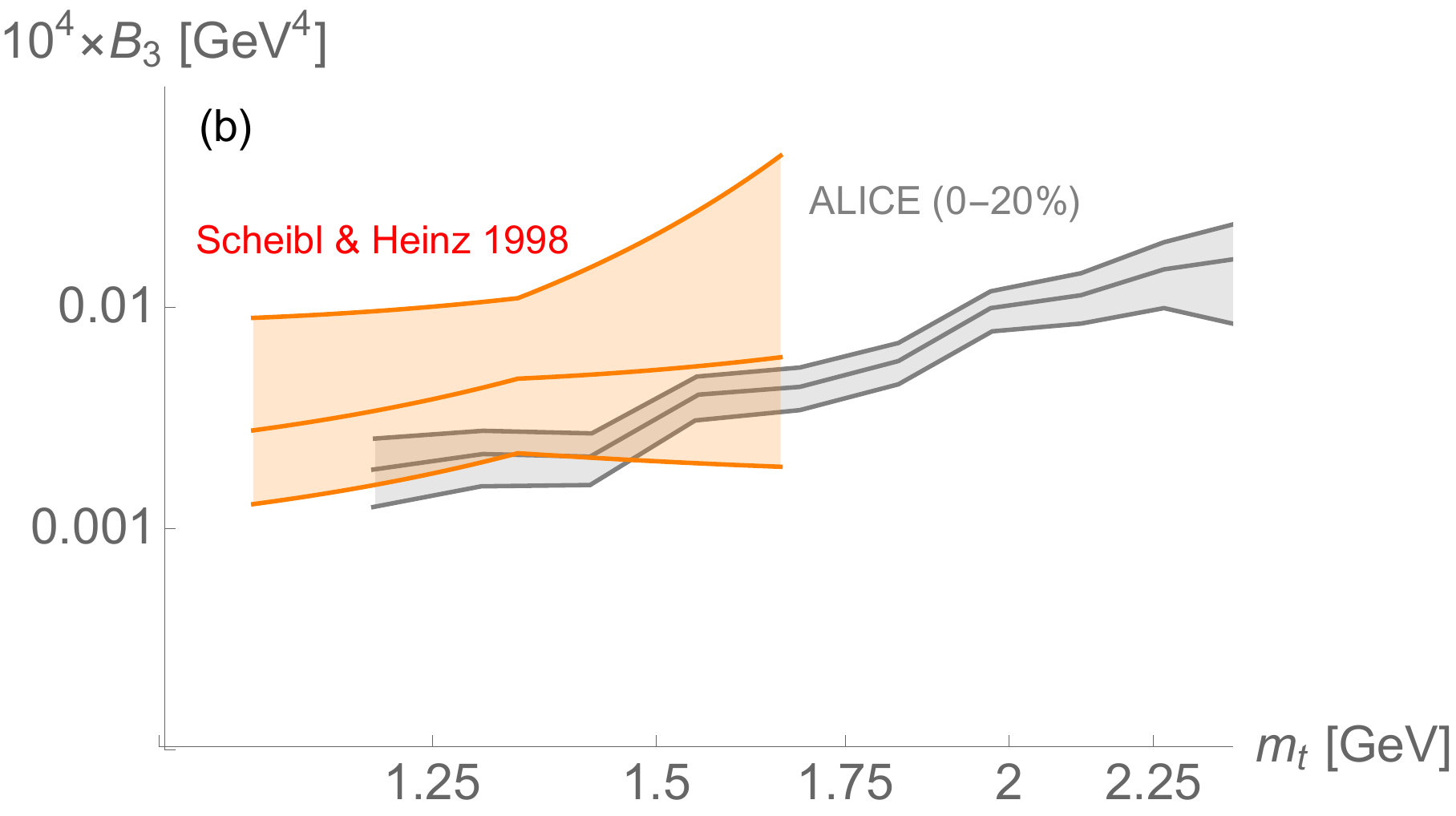}
    \includegraphics[width=0.495\textwidth]{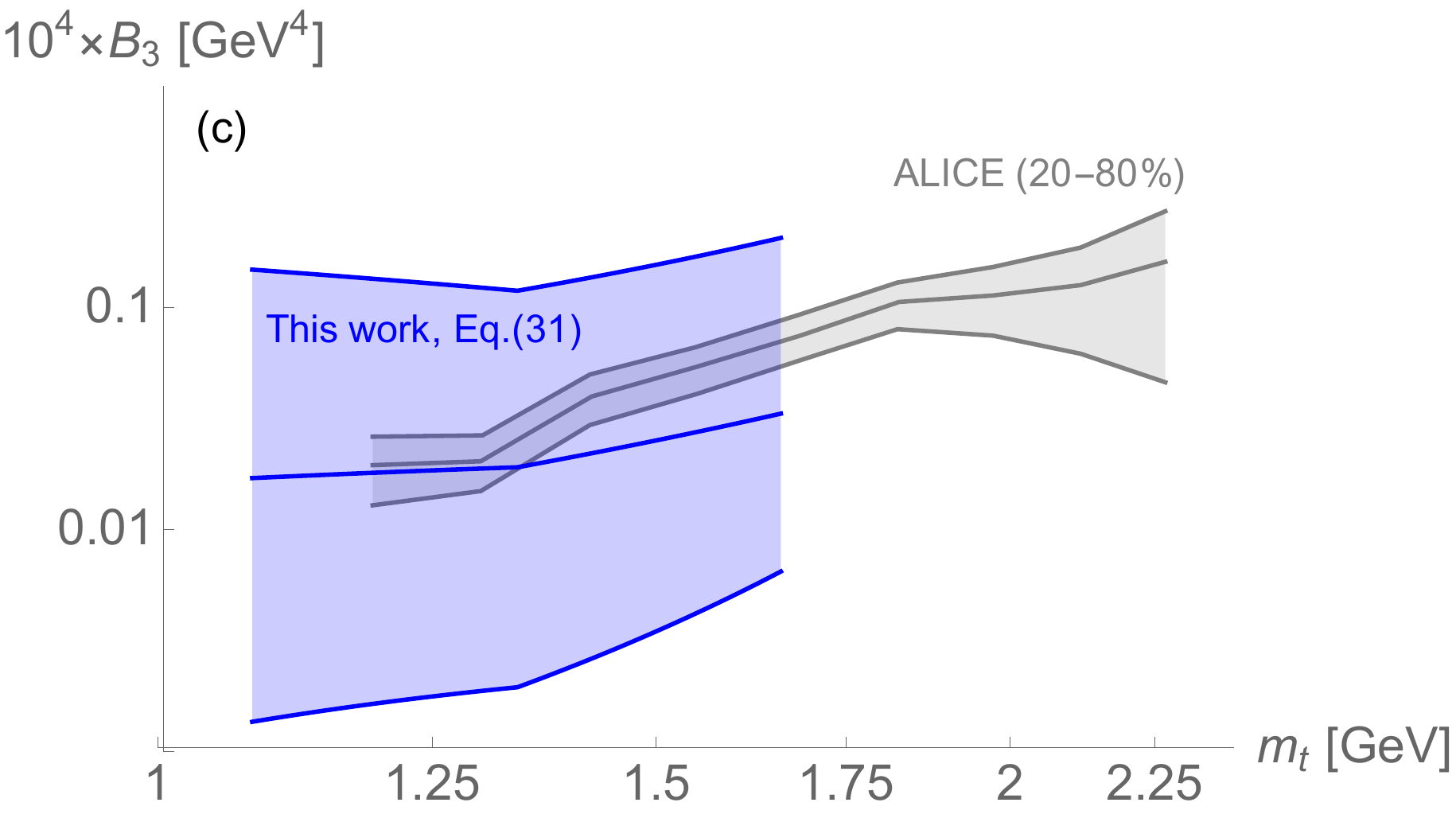}
    \includegraphics[width=0.495\textwidth]{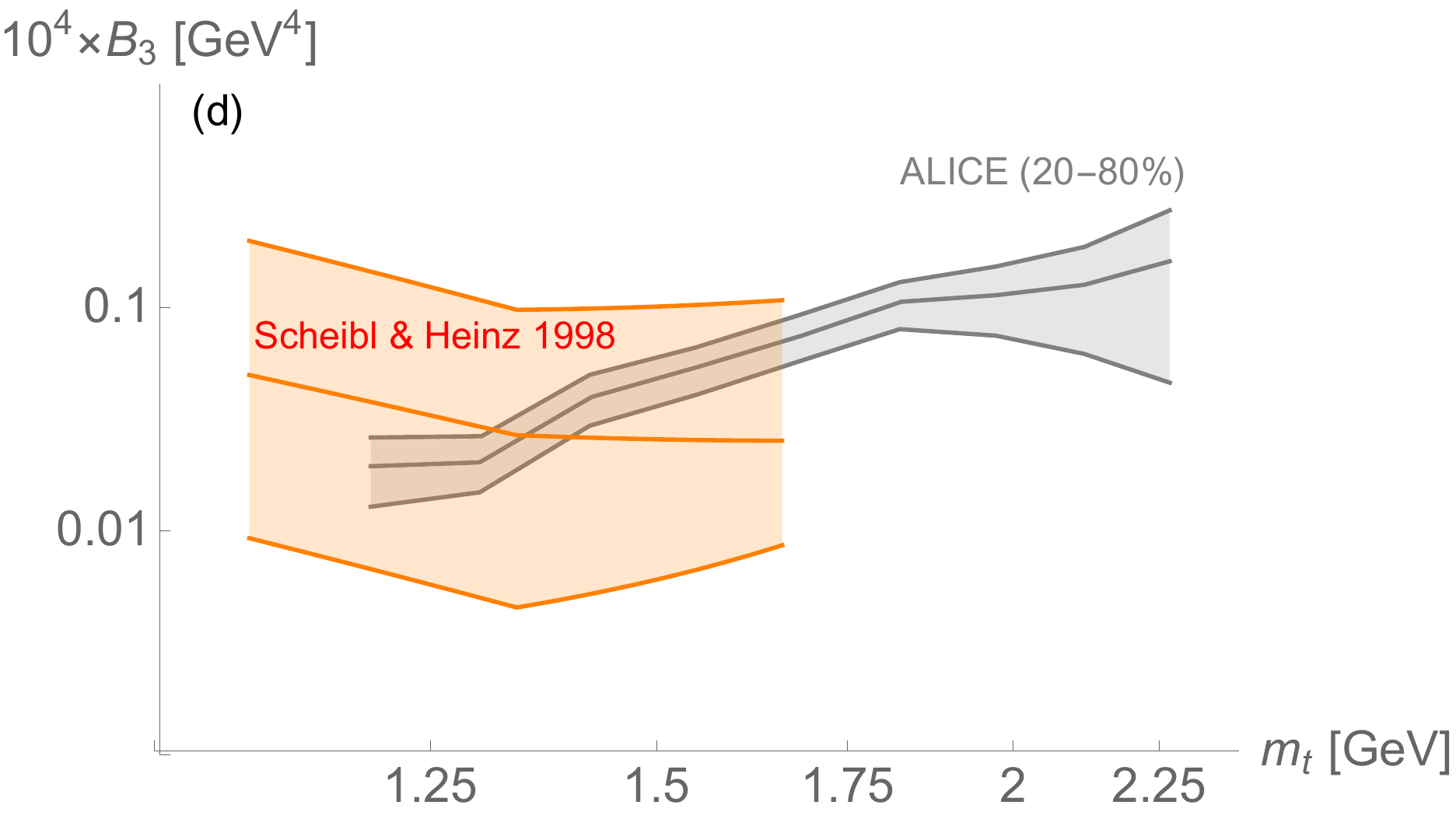}
    \end{center}
  \caption{{\bf Panels (a) and (b):} Experimental results for $\mathcal{B}_3$ from central (0-20\%) PbPb collisions at $\sqrt{s}=2.76$~TeV~\cite{Adam:2015vda}, shown by grey band, compared to Eq.~(\ref{eq:BA}) derived here (blue band) and to the prediction of Ref.~\cite{Scheibl:1998tk} (orange band). {\bf Panels (c) and (d):} experimental values of $\mathcal{B}_3$ from the centrality class (20-80\%), and the theoretical prediction calculated using HBT data from events at (30-50\%).}
  \label{fig:B3PbPb}
\end{figure}

In Fig.~\ref{fig:B3PbPb} we consider experimental results for $\mathcal{B}_3$~\cite{Adam:2015vda} from centrality classes (0-20\%) and (20-80\%), shown in the {\bf top two} and {\bf bottom two panels}, respectively.
%

\subsubsection{p-p collisions}
Ref.~\cite{Acharya:2018gyz} reported $R\approx1.14^{+0.07}_{-0.02}$~fm (comparable statistic and systematic uncertainties were added in quadrature) in a combined analysis of $pp$, $p\Lambda$, and other hyperon correlation data from $\sqrt{s}=7$~TeV p-p collisions at pair average momentum corresponding to $m_t=(1.2-1.6)$~GeV. The analysis in this work effectively assumed $\lambda=1$. However, in an analysis that allowed $\lambda$ to vary as a free parameter, kaon correlations were found to give $\lambda\sim0.5$ at $m_t=1.4$~GeV, along with $R\approx0.8\pm0.3$~fm~\cite{Abelev:2012sq}. This is of potential interest because Ref.~\cite{Adam:2015vja} demonstrated HBT parameters that were the same, within measurement uncertainties, for kaon and proton final states at the same $m_t$.

Using $R\approx1.14^{+0.07}_{-0.02}$~fm as found in the $pp$ analysis~\cite{Acharya:2018gyz}, Eqs.~(\ref{eq:B2SPlam}) predicts $\mathcal{B}_2=10^{-2}\times\left(0.8-0.9\right)\times\lambda~$GeV$^2$. 
Using, instead, $R\approx0.8\pm0.3$~fm as found from kaon correlations~\cite{Abelev:2012sq}, Eq.~(\ref{eq:B2SPlam}) predicts $\mathcal{B}_2=10^{-2}\times\left(0.9-1.4\right)\times\lambda~$GeV$^2$. 
%
These predictions can be compared to light cluster data from Ref.~\cite{Acharya:2017fvb}, which found the experimental result $\mathcal{B}_2^{\rm exp}\approx10^{-2}\times(1.6-2.2)$~GeV$^2$ at $m_t\approx1.4$~GeV.

Using $R\approx1.14^{+0.07}_{-0.02}$~fm~\cite{Acharya:2018gyz}, Eq.~(\ref{eq:BA}) predicts $\mathcal{B}_3=10^{-4}\times\left(2.1-2.8\right)\times\lambda^{\frac{3}{2}}~$GeV$^4$. For $R\approx0.8\pm0.3$~fm~\cite{Abelev:2012sq}, Eq.~(\ref{eq:BA}) predicts $\mathcal{B}_3=10^{-4}\times\left(3.1-23\right)\times\lambda^{\frac{3}{2}}~$GeV$^4$. The experimental result~\cite{Acharya:2017fvb} is $\mathcal{B}_3^{\rm exp}\approx10^{-4}\times(1-3)$~GeV$^4$ at $m_t\approx(1.1-1.4)$~GeV.

\subsection{Discussion: $\mathcal{B}_A$ vs. $R$, coalescence across systems}\label{ss:BAvsR}

Measurement uncertainties on the HBT $R$ and $\lambda$ parameters lead to large uncertainties on our theoretical prediction of $\mathcal{B}_2$ and $\mathcal{B}_3$, derived from Eqs.~(\ref{eq:B2SPlam}-\ref{eq:BA}). Part of this uncertainty is due to our crude treatment of the data. For example, our uncertainty estimate on $\mathcal{B}_{2}$ and $\mathcal{B}_{3}$ in the left panels of Figs.~\ref{fig:B2PbPb}-\ref{fig:B3PbPb} added together the effects of the systematic measurement uncertainties on $R$ and $\lambda$. As a result, while Eqs.~(\ref{eq:B2SPlam}-\ref{eq:BA}) are consistent with the data, there is much room to improve the analysis. 
The coalescence-correlation correspondence motivates an experimental re-assessment of the data presented in Refs.~\cite{Abelev:2012sq,Adam:2015vja,Acharya:2018gyz} and~\cite{Adam:2015vda,Acharya:2017fvb}, aiming at a joint analysis of HBT and cluster yields in events sharing the same $p_t$ and centrality classes.

\begin{figure}[htbp]
  \begin{center}
   \includegraphics[width=0.495\textwidth]{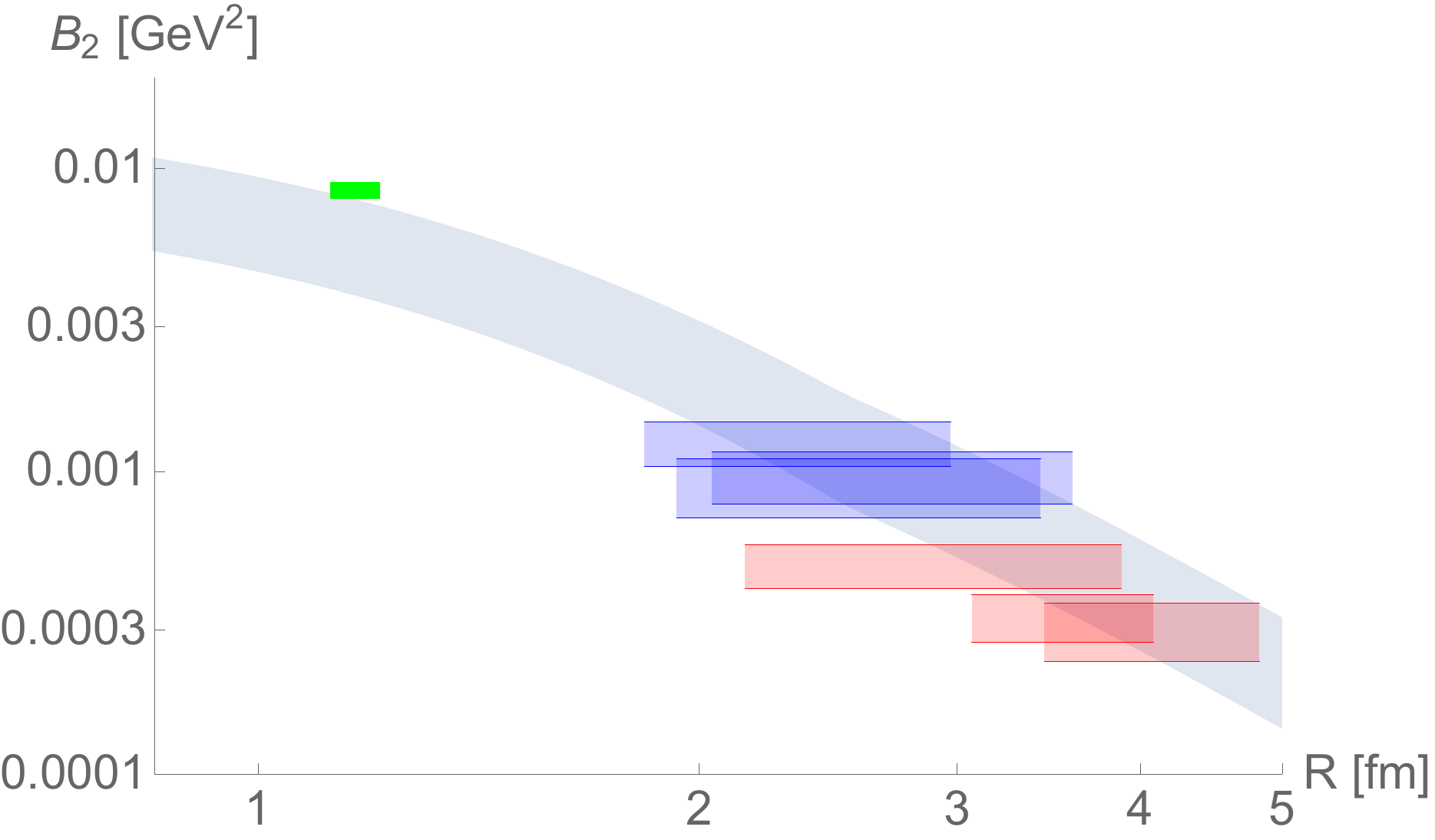}
    \includegraphics[width=0.495\textwidth]{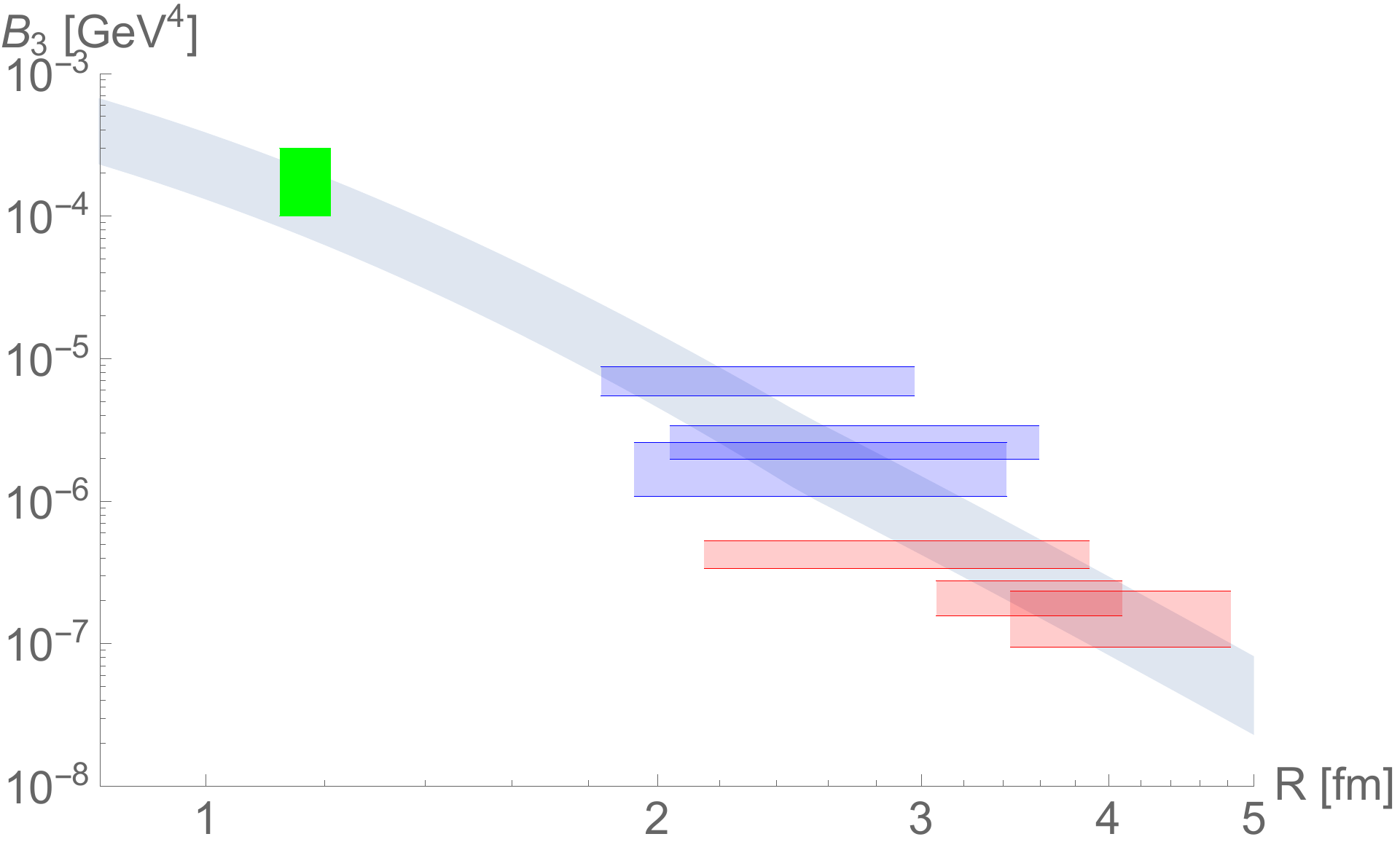}
    \end{center}
  \caption{Summary of data. {\bf Top:} $\mathcal{B}_2$ vs. $R$. {\bf Bottom:} $\mathcal{B}_3$ vs. $R$.}
  \label{fig:BR}
\end{figure}

Before we conclude, in Fig.~\ref{fig:BR} we take a broader look at the data-theory comparison by considering the $\mathcal{B}_A-R$ (anti-)correlation across different systems~\cite{Blum:2017qnn}. In Fig.~\ref{fig:BR}, the grey shaded band shows the theoretical prediction for $\mathcal{B}_2$ ({\bf top}) and $\mathcal{B}_3$ ({\bf bottom}), calculated as function of $R$ using Eqs.~(\ref{eq:B2SPlam}-\ref{eq:BA}). The calculation uses an estimate of the experimentally measured value of $\lambda$. To define the upper edge of the bands, we interpolate between $\lambda=\{1,0.7,0.7\}$ defined at $R=\{0.85,2.5,5\}$. To define the lower edge we interpolate between $\lambda=\{0.5,0.3,0.3\}$ defined at $R=\{0.85,2.5,5\}$. This range of $\lambda$ is roughly consistent with the experimental results found in Ref.~\cite{Adam:2015vja,Acharya:2018gyz,Abelev:2012sq}.
The red horizontal bands in Fig.~\ref{fig:BR} show the (0-10\%) (for $\mathcal{B}_2$) and (0-20\%) (for $\mathcal{B}_3$) coalescence factor measurements for Pb-Pb. Each of the three red bands corresponds to a different bin in $m_t$, among the three bins shown in Ref.~\cite{Adam:2015vja}. The blue horizontal bands show the result for the (20-40\%) (for $\mathcal{B}_2$) and (20-80\%) (for $\mathcal{B}_3$) events, respectively. The green band shows the result for p-p collisions~\cite{Acharya:2017fvb}.

\section{Conclusions}\label{s:sum}

We considered the relation between nuclear cluster formation (defined via a coalescence factor $\mathcal{B}_A$) and two-particle correlation measurements (known as femtoscopy or Hanbury Brown-Twiss (HBT) analyses, with two-particle correlation function $\mathcal{C}_2$) in hadronic collisions. Scheibl \& Heinz~\cite{Scheibl:1998tk} derived a theoretical result, Eq.~(\ref{eq:B2ScheiblHeinz}), equating $\mathcal{B}_A$ to inverse-powers of the source homogeneity radius $R$ measured in HBT analyses. Eq.~(\ref{eq:B2ScheiblHeinz}) is consistent with LHC data over several orders of magnitude in $\mathcal{B}_A$, albeit with large uncertainties~\cite{Blum:2017qnn}. Ref.~\cite{Scheibl:1998tk} based their derivation of Eq.~(\ref{eq:B2ScheiblHeinz}) on a specific, simplified model of collective flow. This model is unlikely to actually represent in detail the dynamics in different systems ranging from Pb-Pb to p-p. The question we addressed to ourselves was, therefore: why does Eq.~(\ref{eq:B2ScheiblHeinz}) work?

Using an idealised quantum mechanical (QM) framework, we derived a direct integral relation between the coalescence factor and the two-particle correlation function. Our main result is Eq.~(\ref{eq:B2C2}), which gives $\mathcal{B}_2$ as an integral of $\mathcal{C}_2$ weighted by the D probability density. The derivation does not require a detailed model of the particle emission source. In particular, we need not invoke the assumptions and approximations of~\cite{Scheibl:1998tk}. If we specialise to the assumptions in~\cite{Scheibl:1998tk}, our  formula essentially reproduces Eq.~(\ref{eq:B2ScheiblHeinz}). Importantly, Eq.~(\ref{eq:B2ScheiblHeinz}) also obtains under more general circumstances if the two-particle correlation function can be approximately described empirically by a Gaussian form, as commonly used in experimental HBT studies.

While our theoretical results are consistent with currently available measurements, the uncertainties are large. Existing experimental analyses were not geared for a direct comparison of femtoscopy and cluster yields. No HBT analysis precisely overlaps, in terms of, e.g., $p_t$ and centrality binning, with cluster yield measurements. The recent study in~\cite{Bellini:2018epz} (see also~\cite{Sun:2018mqq}) proposed to bypass this gap by replacing the HBT part in the coalescence-correlation comparison with multiplicity measurements that correlate with the HBT scales. We suggest, instead, that the coalescence-correlation relation offers a fundamental probe of the (generally defined) coalescence model, justifying dedicated experimental work aiming to test the relation directly.

\acknowledgments
We thank Francesca Bellini, Alexander Kalweit and Urs Wiedemann for discussions and JinJin Pan and Kenny Ng for ongoing collaboration on data analysis related to this work. We are grateful to Ulrich Heinz for discussions and especially for helping us find our way in the literature on coalescence and HBT in heavy-ion collisions. Finally, we thank Bhawani Singh for a diligent reading of our paper and for pointing out a normalisation error. 
KB is incumbent of the Dewey David Stone and Harry Levine career development chair. The work of KB and MT was supported by grant 1937/12 from the I-CORE program of the Planning and Budgeting Committee and the Israel Science Foundation and by grant 1507/16 from the Israel Science Foundation.

\begin{appendix}

\section{Coalescence from correlation functions: kinetic theory}\label{app:kinetic}

Here we give another derivation of Eq.~(\ref{eq:B2C2}). The starting point of our analysis is equivalent to Eq.~(3.12) of Ref.~\cite{Scheibl:1998tk}, derived in Ref.~\cite{Danielewicz:1992pei}.

We assume that the 2-particle source can be factorised as a product of 1-particle source terms. 
The production rate of deuterons (D) at momentum $P_d$, per four dimensional volume in the source region parametrised by D formation coordinates $R$, is given by
\be\label{eq:a0}
         \frac{d}{d^4R}\frac{dN_d}{d^3P_d}
         &=&\frac{3\cdot 2}{(2\pi)^3}
        \int
         \frac{d^3 r\,d^3 Q}{(2\pi)^3}\,
         \mathcal{D}_d\left(\vec{Q},\vec{r}\right)\\
         && f(R_+;Q_+)\,\Gamma_{\rm free}(R_-;Q_-^*), \no
\ea
where the factor $3$ is due to the deuteron spin and the factor $2$
is due to exchange of proton and neutron. $\Gamma_{\rm free}$ indicates the production rate of free nucleons. We have $Q_++Q_-^*=P_d$, and we take $Q_-^*$ slightly off-shell to ensure momentum conservation. 
For small $\vec Q$, we can approximate
\be\label{eq:app1}
         \frac{d}{d^4R}\frac{dN_d}{d^3P_d}
         &\approx&\frac{3\cdot 2}{(2\pi)^3}\int{d^3 r}\,|\phi_d(\vec{r})|^2\\
         && f(R_+;P_d/2)\,\Gamma_{\rm free}(R_-;P_d/2). \no
\ee

It is convenient to consider the coalescence problem in the D rest frame (DRF). In the DRF, we define the source function $S$ as
\be S(x)&=&\frac{m\,\Gamma_{\rm free}(x)}{(2\pi)^3},\ee
such that the free nucleon distribution function is given by
\be
         f(y)&=&\frac{(2\pi)^3}{m}\int_{-\infty}^{y_0} dt\, S(t,\vec{y}).
\ee
For small $|\vec Q|^2\ll m^2$, the constituent nuclei energies are $\approx m$ 
in the DRF, so the Lorentz invariant D yield is 
\be
         \left(E_d\frac{dN_d}{d^3P_d}\right)^{\rm DRF}
         &\approx&
        %
        \frac{2m}{m^2}3\cdot 2(2\pi)^3\int d^4R\,\frac{1}{2}\int{d^4r}\,
        |\phi_d(\vec{r})|^2 \no\\
        &&S\left(R^0-t,\vec{R}-\frac{\vec{r}}{2};m\right)S\left(R^0,\vec{R}+\frac{\vec{r}}{2};m\right).\no\\
        &&\label{eq:a1}
\ee
%

Now, consider the two-point correlation function $\mathcal{C}_2(P,q)$.  
$\mathcal{C}_2(q,P)$ depends on frame and we take the pair centre of mass frame (PRF). For clarity, we use the symbol $\mathcal{C}_2^{\rm PRF}$ to define the two-point function in this frame. 
Under the same source factorisation assumption we considered for the coalescence problem, we have~\cite{Chapman:1995nz}
\be
         \mathcal{C}_2^{\rm PRF}(P,q)&=&\frac{4\int d^4R\int d^4rS\left(R+\frac{r}{2};P\right)S\left(R-\frac{r}{2};P\right)e^{iq\cdot r}}{\left(E\frac{dN}{d^3P}\right)^2},\no\\&&\label{eq:a2}
\ee
where the factor $4$ comes from the spin combinations.

Comparing Eqs.~(\ref{eq:a1}) and~(\ref{eq:a2}), and using Eq.~(\ref{eq:D(k)}), we reproduce Eq.~(\ref{eq:B2C2}).
%
%

\section{Cluster wave function}\label{app:wave}
%
We consider the cluster internal wave function to be a symmetric Gaussian function of the normalised Jacobi coordinates $\vec\xi_n$, $n=1,...,A-1$,
\be\phi_A\left(\vec \xi_1,...,\vec \xi_{A-1}\right)&=&\frac{\exp\left(-\frac{\sum_{i=1}^{A-1}\vec\xi_i^2}{2d_A^2}\right)}{A^{\frac{3}{4}}\left(\pi \,d_A^2\right)^{\frac{3(A-1)}{4}}},\ee
where~\cite{Shebeko:2006ud}
\be\vec\xi_n&=&\frac{n}{\sqrt{n^2+n}}\left(\vec r_{n+1}-\frac{1}{n}\sum_{m=1}^{n}\vec r_m\right)\ee
and where $\vec r_m$, $m=1,...,A$ are the Cartezian constituent nucleon coordinates. 
The size parameter $d_A$ is related to the cluster rms charge radius via~\cite{Mattiello:1996gq,Scheibl:1998tk,Bellini:2018epz}
\be r^2_{\rm rms}&=&\frac{3(A-1)}{2A}d_A^2.\ee

\end{appendix}

\vspace{6 pt}

\bibliography{ref}

\end{document}